# EchoAid: Enhancing Livestream Shopping Accessibility for the DHH Community

ZEYU YANG, The Hong Kong University of Science and Technology (Guangzhou), China
ZHENG WEI, The Hong Kong University of Science and Technology, China
YANG ZHANG, Tencent, China
XIAN XU, Lingnan University, China
CHANGYANG HE, Max Planck Institute for Security and Privacy, Germany
MUZHI ZHOU*, The Hong Kong University of Science and Technology (Guangzhou), China
PAN HUI†, The Hong Kong University of Science and Technology (Guangzhou), China

Livestream shopping platforms often overlook the accessibility needs of the Deaf and Hard of Hearing (DHH) community, leading to barriers such as information inaccessibility and overload. To tackle these challenges, we developed *EchoAid*, a mobile app designed to improve the livestream shopping experience for DHH users. *EchoAid* utilizes advanced speech-to-text conversion, Rapid Serial Visual Presentation (RSVP) technology, and Large Language Models (LLMs) to simplify the complex information flow in live sales environments. We conducted exploratory studies with eight DHH individuals to identify design needs and iteratively developed the *EchoAid* prototype based on feedback from three participants. We then evaluate the performance of this system in a user study workshop involving 38 DHH participants. Our findings demonstrate the successful design and validation process of *EchoAid*, highlighting its potential to enhance product information extraction, leading to reduced cognitive overload and more engaging and customized shopping experiences for DHH users.

---

*Corresponding authors
†Corresponding authors

---

Authors' Contact Information: Zeyu Yang, zyang979@connect.hkust-gz.cn, The Hong Kong University of Science and Technology (Guangzhou), Guangzhou, China; Zheng Wei, zwei302@connect.ust.hk, The Hong Kong University of Science and Technology, Hong Kong, China; Yang Zhang, e4a_eureka@foxmail.com, Tencent, Guangzhou, China; Xian Xu, xianxu@LN.edu.hk, Lingnan University, Hong Kong, China; Changyang He, Max Planck Institute for Security and Privacy, Bochum, Germany, changyang.he@mpi-sp.org; Muzhi Zhou, mzzhou@hkust-gz.edu.cn, The Hong Kong University of Science and Technology (Guangzhou), Guangzhou, China; Pan Hui, panhui@hkust-gz.edu.cn, The Hong Kong University of Science and Technology (Guangzhou), Guangzhou, China.

---

# 1  INTRODUCTION

Livestreaming shopping platforms have transformed how consumers shop by combining Internet technology with real-time interaction, leading higher conversion rates [75]. The rise of livestream platforms has encouraged more merchants to utilize livestream to showcase their products, usually at a lower price than other online shopping formats. In China, the marriage between livestreaming and e-commerce has attracted over 500 million users and generating a market worth 4.9 trillion RMB in 2023 [55].

The fast-paced and information-heavy nature of livestreaming poses accessibility challenges for the Deaf and Hard of Hearing (DHH) community [10, 11]. Consumers rely on live streaming and real-time information from streamers throughout their decision making process [20]. The ability to exchange information instantly and engage with the streamer in a one-to-many format is crucial for making timely and informed purchase decisions. With limited time available, factors such as price promotions and visual appeal play a significant role in influencing Chinese consumers' impulse buying decisions [34]. Additionally, interactions between consumers and streamers, as well as among consumers themselves, are important and often enjoyable, when livestreams are often entertaining [74]. The simultaneous engagement of consumers in live e-commerce involves a variety of activities [70, 83], such as interacting with streamers, evaluating product features, and purchasing via mobile payment, making the information exchange in online shopping more complex and intensive than typical online conferences. This nearly parallel processing of product information, participants' social information, and economic decision-making significantly increases the information and operational burden for DHH users.

Technologies such as automatic speech recognition (ASR) have been developed to assist communication for DHH users [15, 30, 40]. However, in situations like livestream shopping, where important information is conveyed both visually and audibly, traditional methods such as enhanced captioning and sign language [35], as well as real-time speech-to-text conversion may become less effective. This is because the fast-paced delivery of information, such as the dense text directly translated from speech, can create overwhelming visual stimuli, leading to significant cognitive overload [68] and impairing consumers' ability to make informed purchasing decisions.

Given these challenges, especially the issue of cognitive overload for DHH users during livestream shopping, we introduce *EchoAid*, a mobile application designed to enhance the livestream shopping experience. *EchoAid* aims to address common challenges faced by the DHH community in online shopping, such as information loss, misunderstanding, and information overload [18, 27, 44]. We incorporate customized and condensed speech-converted text in real-time by combining Rapid Serial Visual Presentation (RSVP) technology with Large Language Models (LLMs), where key sales information is highlighted through keyword extraction techniques. *EchoAid* also offers a customized graphical user interface (GUI) with adjustable caption backgrounds. Additionally, *EchoAid* enhances user engagement and accessibility by supporting cross-platform compatibility.

In our study, we first interviewed eight DHH individuals to establish the primary design guidelines. Three of these participants were later involved in prototype design tests and co-design workshops to improve the system. To evaluate *EchoAid*, we conducted a thorough user study with 18 DHH participants to measure its effectiveness in reducing information overload and enhancing livestream shopping experiences by comparing our *EchoAid* with "*iFlytek Hearing*", which is a popular mobile application among DHH that can convert speech into texts. The findings of these experiments and interviews show that *EchoAid* significantly enhances information retention and decreases the cognitive load associated with task management for users. This study was approved by the Institutional Review Board (IRB) of our University, approval No. HSP-2024-0019.

Driven by the motivation to enhance information acquisition for DHH during livestream shopping, our *EchoAid* effectively minimizes information loss and overload that significantly hinder DHH users' active participation in livestream shopping. Taking advantage of the language capabilities of LLM, our design provides a condensed speech-to-text summary of sales information in real time, helping to mitigate confusion arising from common errors made by widely commercialized speech-to-text software such as *iFlytek*. The short text summaries of key sale information enable DHH users to capture information transmitted audibly as well as visual cues related to product information in a limited time. These newly added livestream shopping features in our system have reduced information loss and overload, making stimuli conveyed by the livestreaming host more effective. Furthermore, *EchoAid* features a tailored interface, incorporating customizable graphical user interface (GUI) elements and caption windows to enhance personalization and engagement. Our user studies have overwhelmingly supported the effectiveness of *EchoAid* in facilitating more active and accessible participation in livestream shopping. We believe that with advancements in speech-to-text technologies, *EchoAid* has great potential to enhance the online shopping experience for DHH users, allowing them to enjoy the benefits of a more interactive, entertaining, and authentic livestream shopping experience.

## 2 BACKGROUND AND RELATED WORK

We begin with a brief introduction to livestream shopping, followed by an identification of two main challenges faced by DHH consumers in this context: access to product information (resulting in information loss) and information overload, which consequently results in diminished enjoyment of the viewing experience and even withdrawal from livestream shopping.

### 2.1 Live e-commerce and livestream shopping

Live e-commerce, known as livestream shopping, is a retail concept that combines traditional online shopping with real-time interactive video content. Originating in China, livestream shopping has become popular on platforms like *Taobao* and *Douyin* [1], and has attracted increasing research attention in HCI and CSCW community [47, 70, 77, 83]. During livestream sessions, hosts, often influencers or sellers, showcase products, explain their features, and interact with viewers (usually by viewing viewers' comments) in real-time, creating an engaging shopping experience [66, 74, 82]. The interactive nature of these sessions boosts viewers' confidence in the products and purchasing desire to purchase [54, 67]. Hosts can offer products at discounted prices by utilizing bulk purchases and direct sales from manufacturers, often providing better deals than traditional retail stores both offline and online [48, 88].

Livestream shopping is widespread on social media platforms in China, but there is a lack of tailored accessibility features for the DHH community. Existing smartphone accessibility tools, such as voice-to-text functions, may not accurately transcribe speech at the fast pace often seen in livestream product presentations [10]. This can result in delayed subtitles and information overload for DHH users, leading to unpleasant viewing and shopping experiences [16].

### 2.2 Information Loss and Overload of DHH Users in Livestream Shopping

We use the Stimulus-Organism-Response (S-O-R) model to identify key challenges faced by DHH consumers in the context of livestream shopping. S-O-R model is a common framework for analyzing consumer behavior [21, 46]. As shown in Figure 1, in livestream e-commerce, key stimuli include the host's personal charisma, real-time interactions, dynamic visual effects, and limited-time promotions (S) [45]. Shopping information is reflected via the auditory cues, visual elements, and interactive features of the live stream. The hosts in livestream shopping usually speaks very fast to create a sense of urgency. The reliance on auditory information for shopping knowledge creates a significant barrier for DHH users to accessing critical product details and promotional content. The absence of effective real-time captioning or sign language interpretation means that DHH viewers may miss out on essential information, which is referred to as **information loss**. This lack of access to stimuli can hinder their ability to make informed purchasing decisions.

Even when DHH consumers do manage to access information, they often face the challenge of **information overload**. Information overload occurs when an individual is exposed to an excessive amount of information that surpasses their processing capabilities. This issue is prevalent when there are multiple information sources, rapid information dissemination, and exponential growth in data volume [4]. This overload can result in higher cognitive load, difficulties in decision-making, fragmented attention, and increased psychological stress [19]. Research has shown that managing large amounts of text and real-time conversion and display of information pose additional complexities for DHH individuals [10].

Moreover, the S-O-R model suggests that consumers' flow experience and the reduction in the perceived risks (O) can be enhanced through interpersonal interactions during live streams [33, 74]. The advancement of live commerce lies in the ability of live streaming to achieve real-time interaction with fans. When fans have questions, they can directly leave messages in the live stream, and the host can interact with them through the screen, providing online assistance and answers. This aspect restores the scene of offline shopping. For DHH viewers, the rapid delivery of content that is transcribed to text—combined with dense visual stimuli—further amplify the significance of information overload. Their cognitive resources are further taxed by trying to keep up with the flow of all different types of visual cues in a fast-paced manner. This overload can result in confusion and frustration, further impeding their ability to engage effectively with the shopping experience.

Those combined stimuli and organism factors lead consumers to exhibit purchase intentions, which can translate into actual buying behavior, which is the response (R) [14]. For DHH viewers, the culmination of these challenges–limited access to information and the burden of cognitive overload–leads to diminished enjoyment of the livestream shopping experiences, further affecting their willingness to participate in future livestream shopping events, as negative experiences can lead to disengagement from this online shopping format.

The S-O-R model directs us to identify the key barriers DHH consumers face in livestream shopping. It also serves as an instructive framework guiding our later interview with DHH users and system design goals. In particular, we focus on information loss and overload issues, recognizing them as critical bottlenecks that impair comprehension, decision-making, and overall shopping satisfaction. These insights from the S-O-R model directly shaped our interview directions to gather first-hand perspectives from DHH users to confirm these insights drawn from the S-O-R model. They later guided our user study evaluation to assess information comprehension and shopping satisfaction.

## 2.3 Advancements in Translating Audio to Text and Visual Cues

Various speech-to-text technologies have been developed to enhance general accessibility for the DHH community [24, 44]. For instance, "*iFlytek* Hearing" offers a mobile application that converts spoken words into text messages, facilitating real-time communication. Additionally, *Voibook*[1], a

---

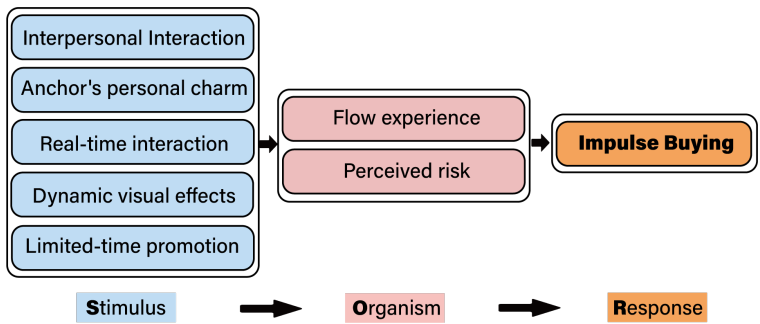

Fig. 1. The S-O-R (Stimulus-Organism-Response) model in livestream shopping illustrates how factors like interpersonal interaction, anchor charm, real-time engagement, dynamic visuals, and limited-time promotions influence consumers' psychological states (e.g., enhanced flow experience and reduced perceived risk), ultimately driving impulse buying behavior.

specialized communication tool for the DHH community, supports both speech-to-text and text-to-speech conversions, enabling bidirectional communication. *Huawei* has also introduced an AI-powered smart subtitle feature in their smartphones[2], which can transcribe speech from videos or live conversations into subtitles onscreen, including real-time translation into Chinese.

The translation of audio input is not limited to text. Michael V. Perrotta proposed multi-sensory feedback devices through visual, tactile and auditory feedback [57]. AR technology has also been employed to translate audio input into both text and visual cues and support DHH users in educational and interactive contexts, demonstrating its ability to support better academic engagement and outcomes [30, 36, 60, 81, 81].

To date, there has been limited research focused on enhancing the accessibility of livestream shopping platforms specifically for the DHH community. While existing general speech-to-text technologies, such as real-time subtitles, have been developed, they often fall short in addressing the unique communication barriers faced by DHH users in a livestreaming context. Unlike typical live conference communication, livestream shopping combines rapid speech delivery with extensive visual cues related to product sales. This dynamic environment demands more effective accessibility solutions that can accommodate both the auditory and visual elements simultaneously. The existing research gap highlights the urgent need for inclusive technologies that ensure all user groups, including the DHH community, can fully benefit from the conveniences offered by livestream shopping.

## 2.4 RSVP to address information loss

Despite recent technological advancements, the specific needs of DHH individuals during livestream shopping have not been adequately met. Existing software solutions do not fully account for the challenges posed by livestream environments, where rapid speech in combination with visual cues of the product under sale, as well as background noise can impair the effectiveness of speech recognition systems.

In such cases, the quick presentation of text information is crucial. Methods to speed up access to information while preserving the key meaning include: moving windows, time squares, line stepping, sentence-by-sentence presentation, and rapid serial visual presentation or RSVP [12].

---

[2]https://consumer.huawei.com/cn/support/content/zh-cn15812021/

RSVP was originally considered an experimental paradigm for studying attentional mechanisms. Gilbert [26] first proposed it in a reading context in the late 1950s, while Forster [25] used it to study the comprehension and processing of written language. RSVP refers to the sequential presentation of one or more words at a time, thereby minimizing the eye movements generated during reading and improving concentration.

The development of RSVP technology has proven effective in enhancing attention and reading abilities among the DHH community [61, 72]. Unlike traditional text-to-speech technologies, RSVP presents information continuously and rapidly in a word-by-word or letter-by-letter format. This form of information presentation has the potential to offer a clear and smooth reading experience even in the fast-paced and noisy setting of livestream broadcasts. Used in real-time AR caption interface in DHH students' classrooms, text display digital radio, and television multimedia subtitling, RSVP excels in addressing challenges related to background noise and fast speech, thereby ensuring that DHH users receive accurate and understandable information [27, 62, 65].

## 2.5 LLM summary to address information overload

Existing methods to assist DHH individuals in acquiring information, leveraging advanced speech-to-text translation technologies may contribute to information overload [2, 4, 52]. As have already mentioned, this issue is even more severe in livestreaming shopping. To address information overload, researchers have proposed strategies such as keyword extraction, highlighting only key information, in combination with RSVP technology [71]. These solutions rely on precise text inputs to ensure the accuracy and speed of information processing.

Inspired by the idea of key information extraction, we propose the use of LLMs to summarize speech-to-text information effectively, designating the LLM to condense and present key sale information in a more structured way. By combining LLMs with RSVP display techniques, we aim to refine and emphasize essential information during livestreaming shopping. We hope this approach can empower DHH users to engage more efficiently in real-time communications and interactive activities.

## 3 STUDY I: UNDERSTANDING THE LIVESTREAM SHOPPING EXPERIENCE OF DHH INDIVIDUALS

### 3.1 Method

In this study, we conducted semi-structured interviews with eight participants to explore the user experiences of DHH individuals in livestream shopping. Our objective was to identify the key issues and needs of the DHH community related to their use of livestream shopping platforms, focusing on viewing habits, interaction experiences, technical barriers, and prevalent challenges.

Participants were recruited through online social media, provincial Disabled Persons' Federation, and NGOs. Selection criteria included self-reported hearing loss and regular use of livestream shopping platforms for at least 2 years with monthly purchases on these platforms.

The demographics of the participants are detailed in Table 1. There were five males and three females (Age $\bar{M}$ = 37.125, SD = 9.96) with a wide range of hearing disabilities. Interviews were conducted with professional sign language interpreters, and recordings were transcribed. Interviewing questions are listed in Appendix A.3. Each interview session lasted approximately 30 minutes, and participants received compensation of 20 RMB for their time.

Thematic analysis was performed using open coding. Initially, all researchers reviewed the entire transcript to gain a basic understanding and identify any errors. After that, two researchers watched the recorded video and checked with the sign language interpreter again to correct any transcription mistakes. They also independently examined the transcribed texts and created the first level of coding based on recurring keywords and content similarity. Any differences in their findings were discussed and resolved during weekly meetings with a senior researcher, which led to an agreement on the sub-themes for each major theme that are aligned with the study's objectives. Ultimately, all researchers convened and decided on four main themes. This thematic analysis, utilizing inductive coding methods, adheres to established qualitative research guidelines [69].

Table 1. **Study I: Demographics of 8 DHH Participants. DHH State: Mild(speech audible with difficulty in subtle sounds), Moderate(loud or close-up speech needed), Moderately severe(dialogue inaudible, only very loud sounds heard), Severe(near-total deafness, requires aids or implants), Totally deaf(no hearing). Livestreaming Usage(How often do you usually watch content on the live streaming platform): Daily, Weekly, Monthly, Occasionally, Never. Assistive Tech: The most commonly used software for watching live streaming platforms.**

| ID | Gender | Age | DHH State | Sign Language Usage | Livestreaming Usage | Assistive Tech |
|----|--------|-----|-----------|---------------------|---------------------|----------------|
| P1 | M | 51 | Moderate | Occasionally | Weekly | *iFlytek* , *Voibook* |
| P2 | F | 37 | Severe | Frequently | Weekly | *iFlytek* , *Voibook* |
| P3 | M | 52 | Totally deaf | Frequently | Weekly | *iFlytek* , *Voibook* |
| P4 | F | 34 | Severe | Always | Daily | *iFlytek* |
| P5 | M | 24 | Severe | Never | Occasionally | *iFlytek* |
| P6 | F | 25 | Moderately severe | Frequently | Occasionally | *iFlytek* |
| P7 | M | 42 | Moderately severe | Never | Occasionally | *iFlytek* |
| P8 | M | 32 | Severe | Frequently | Weekly | *iFlytek* , *Hua wei Ai* , *Voibook* |

## 3.2 Main Findings

All participants indicated a strong interest in watching livestream shopping regularly, citing the availability of freebies and competitive pricing compared to other shopping avenues.

The four primary challenges that impact their livestream shopping platform experience are:

**Operational Difficulty**

Participants relied on multiple devices to capture livestreaming content, primarily stemming from the absence of real-time subtitles and sign language translation on those platforms. Switching between those devices posed extra operational challenges. User 5 said, *"I require two old phones for livestream shopping; one for viewing and another for translating on the side, but managing both at the same time is daunting."*

Moreover, the fast-paced nature of livestream shopping, involving multiple hosts or assistants, complicates voice recognition and speaker identification. The lag in speech-to-text translation substantially hindered our participants' ability to follow and engage. Users 3 and 7 expressed frustrations with interacting through comments, noting, *"When I use comments to interact with the host, I often don't get a response,"* and *"The experience is frequently frustrating due to delays and inaccuracies with the translation software I use."* User 4 explained, *"Current systems for multicharacter chat and translation fall short of our needs. A text refinement tool could be of great help to us deaf and mute individuals."* These observations underscore the pressing need for more efficient and user-friendly interaction mechanisms on livestream shopping platforms to enhance the participatory experience of DHH users.

**Inaccurate Voice Recognition** Participants highlighted issues related to nonrecognition or misunderstanding with current voice recognition and translation technologies during their use. Common problems include low accuracy in voice recognition (mentioned by users 1, 2, 3, 5, 7) and difficulties in recognizing accents and dialects (mentioned by users 3, 5, 7, 8). For instance, User 7 expressed frustration, stating, *"The subtitles from the iFlytek Voiceair I use often misinterpret words or phrases; I can't understand the livestream content or participate in time, causing me to miss out limited-time offers."* Additionally, User 4 highlighted, *"When multiple people are speaking in a livestream* *shopping room, the translation software fails to distinguish between the speakers, making it confusing for me to follow.*" This lack of accuracy significantly hinders DHH users from comprehending the content fully, emphasizing the necessity for more robust and precise voice recognition solutions.

**Information Overload** All participants raised concerns about experiencing information overload while using livestream shopping platforms. They expressed difficulties with voice-to-text translation software that presented excessive text within a short timeframe, making it challenging for them to process and retain essential information. User 6 stated, *"I am really afraid of handling too much text in a short time. This increases my cognitive load to the point of mental shutdown."* User 2 also voiced frustrations, stating, *"When watching a livestream shopping, the host speaks too fast, and I have to watch both the iFlytek subtitles and the live comments, which move quickly and are in small font, making it hard to process all this information at once."* User 8 emphasized, *"It is really important to simplify information to prevent overwhelming the brain."* User 8 emphasized the importance of simplifying information to avoid overwhelming the brain. These insights underscore the need for a system that can streamline and simplify information presentation while highlighting critical sale content to alleviate cognitive strain.

**Unattractive Interaction Interfaces**

Participants also linked their reduced engagement with livestream shopping to the unappealing and uninspiring user interfaces commonly found in many livestream shopping platforms. Multiple users expressed dissatisfaction with the static and text-centric interfaces, noting that these designs fail to captivate their interest, resulting in a lackluster shopping experience. User 4 remarked, *"The interface is dull and repetitive; it feels like I'm merely observing a continuous text stream, which does not foster active participation."* User 5 echoed similar sentiments, stating, *"The absence of visual and interactive elements renders the livestreams unexciting and tedious, often discouraging me from sustained engagement."* This feedback underscores the imperative for livestream platforms to integrate more dynamic and visually appealing features to engage and retain users' attention, particularly those within the DHH community who heavily rely on visual cues.

## 4 STUDY II: ECHOAID DEVELOPMENT

### 4.1 Design

Through a synthesis of information from the S-O-R model literature, the system-related studies and user interviews, we identified six key design considerations, as summarized in Figure 2:

**D1: Support in-device cross-software voice-to-text.** DHH individuals typically interact with livestream shopping platforms by initiating a speech-to-text translation application, such as *iFlytek*, set to overlay mode, before selecting a livestream to view. Additionally, some DHH users, constrained by outdated devices or financial limitations, resort to using multiple devices to engage in livestreams, thereby complicating their ability to multitask effectively.

**D2: Support key sale information extraction and presentation.** During livestream shopping, hosts often discuss a mix of relevant and irrelevant details during product promotions, leading to information complexity, overload, and the risk of key sale information loss [30, 58] and increase information overload and the risk of information loss. Existing solutions use keyword extraction and text summarization techniques to filter out irrelevant content but may lack accuracy and fail to refine information effectively [7, 87].

In developing the *EchoAid* system, we compared speech-to-text applications like *YINSHU*, *iFlytek*, and *Huawei*, incorporating feedback from DHH user interviews. Current tools efficiently transcribe speech but produce verbose outputs, requiring users to filter through excessive information. While some tools extract key phrases, they may present fragmented or inaccurate contexts, adding to users' cognitive load.

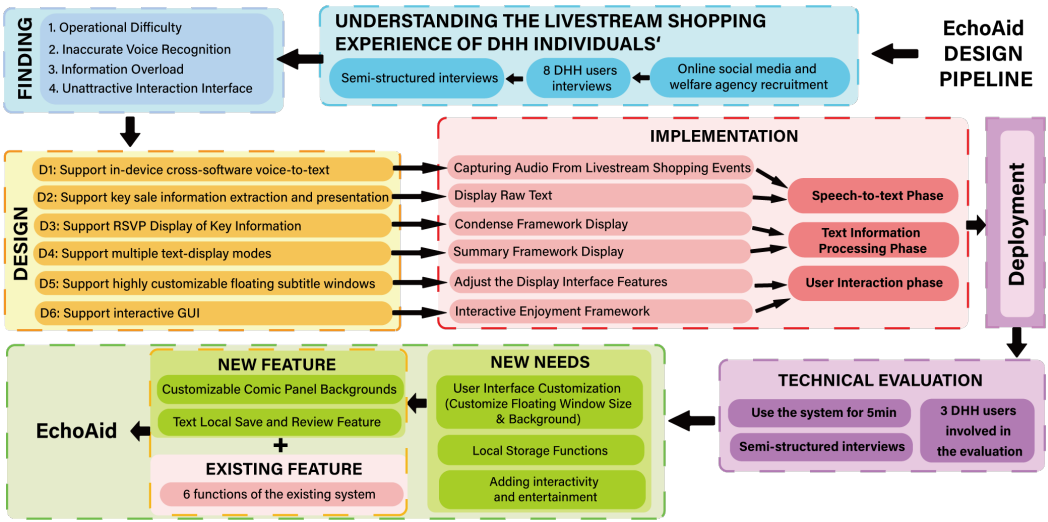

Fig. 2. Research design. Initial interviews identified four key challenges. Based on these findings, six design solutions were proposed. These designs guided the development and deployment process, summarized into three phases: speech-to-text, text processing, and user interaction. After deployment, three DHH users tested the system and participated in interviews and co-design sessions, uncovering three new needs. To address these needs, two additional features were developed. This process resulted in eight new features.

Accordingly, one of the key functions of *EchoAid* should be able to significantly reduce the volume of information presented to DHH users, focusing on displaying only the most relevant sale details.

**D3: Support RSVP Display of Key Information.** RSVP technology presents text sequentially, either word by word or letter by letter, at a controlled speed tailored to users' processing abilities. This method aids in managing information flow efficiently, lowering cognitive load by delivering text in a more digestible format. It enables DHH users to stay abreast of rapid information during livestream shopping without overlooking essential details. By prioritizing the optimization of presentation speed and sequence, the *EchoAid* system strives to refine information processing and enhance the accessibility of livestream shopping for the DHH community.

**D4: Support multiple text-display modes.** Considering the preferences of DHH users and designing an accessibility-friendly user interface is crucial to enhance their engagement and user experience [28, 42, 50, 64]. Users engage in livestream shopping for various reasons, such as purchasing specific products, browsing for ideas, or simply passing time. Therefore, it is essential to provide DHH users with the option to choose their preferred text display mode.

In the design of *EchoAid*, we incorporated three text display modes: Raw Text Display, Condensed Text Display, and Summary Framework Display, as illustrated in Figure 6 (1)-(3). These modes, coupled with efficient keyword extraction technology, enable users to quickly grasp key information during livestream, significantly enhancing their overall viewing and shopping experience.

**D5: Support highly customizable floating subtitle windows.** During livestream shopping, it is essential to be able to adjust the position and size of floating windows as hosts or products may change their sizes and positions on the screen. Current speech-to-text tools, like the *iFlytek* software on *iOS*, only permit proportional scaling and restrict the placement of caption windows to the four corners, limiting customization options. On *Android* systems, while custom positioning is feasible, resizing is not supported. Only *Huawei*'s official AI subtitle function offers both custom size and position adjustments, but this feature is exclusive to specific mobile phone models. *EchoAid* breaks through these limitations, not only supporting the position and size adjustment of floating windows, but also allowing users to customize window transparency and backgrounds with simple swipe gestures (Figure 5 (2)-(2-1)). After the settings are complete, users can easily switch between different interface modes (Figure7 (2)-(2-2)) to create a highly personalized live streaming viewing experience based on their individual needs.

**D6: Support interactive GUI.** The *EchoAid* system has integrated a livestream graphical user interface (GUI) interactive dialogue interface to enhance interactive engagement during livestream. This design not only improves the intuitiveness of information transmission but also enhances user engagement and experience. For instance, when the host discusses product features, corresponding icons are displayed on the interface (Figure7-(4-1)), updating in synchronization with the content in the RSVP box. Different GUI icons represent various information categories. Using these live GUI elements, the *EchoAid* system can present complex information in a simple and intuitive manner, enhancing readability and user engagement, and enriching the user's interactive experience.

## 4.2 Implementation

*EchoAid* is a smartphone application designed to improve the livestream shopping experience for the DHH community. Figure 3 outlines the components of the *EchoAid* system, and the development code can be accessed on GitHub at https://anonymous.4open.science/r/echoaid_1007/readme.md.

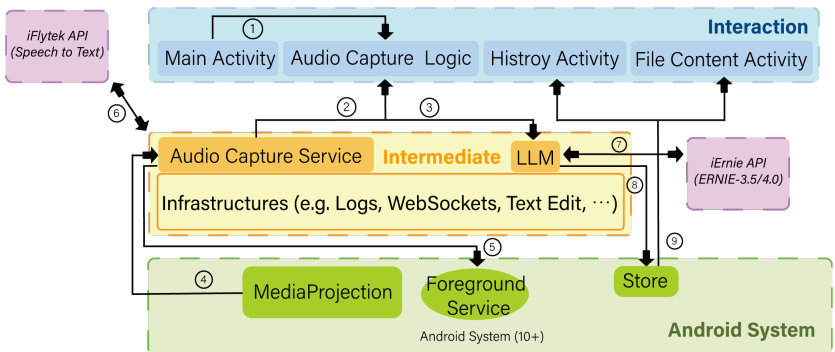

Fig. 3. System architecture. Main Activity (①) provides the main interface, allowing users to adjust the appearance of Audio Capture Logic and access activity history. Audio Capture Logic (②, ③) displays captions, offers toggle buttons, and clears context, while controlling both the Audio Capture Service (②) and LLM (③). In the middle layer, Audio Capture Service (④, ⑤, ⑥) captures MediaProjection from the system via API (④), registers as an Android foreground service (⑤), and uploads the real-time audio stream to the iFlytek API for speech-to-text conversion (⑥). Meanwhile, LLM (⑦, ⑧) handles large model interactions by generating summaries, frameworks, and emojis (⑦) and calling external services when needed (⑧). Finally, the generated results are saved in system storage and accessed by History Activity and File Content Activity (⑨).

### 4.2.1 Speech-to-text Phase.

The initial phase of *EchoAid* involves capturing audio from livestream shopping events and converting it to text. This is accomplished using the iFlytek MediaProjection API for cross-program audio capture (D1). *EchoAid* submits 640-byte PCM audio every 40 milliseconds to the real-time *iFlytek* API via WebSocket, which processes the audio for its precision and speed and returns the speech-to-text result. As shown in Figure 7 (2)-(3), audio is processed through cloud servers, and the resulting text is presented on the *EchoAid* interface, offering various user-customizable display modes.

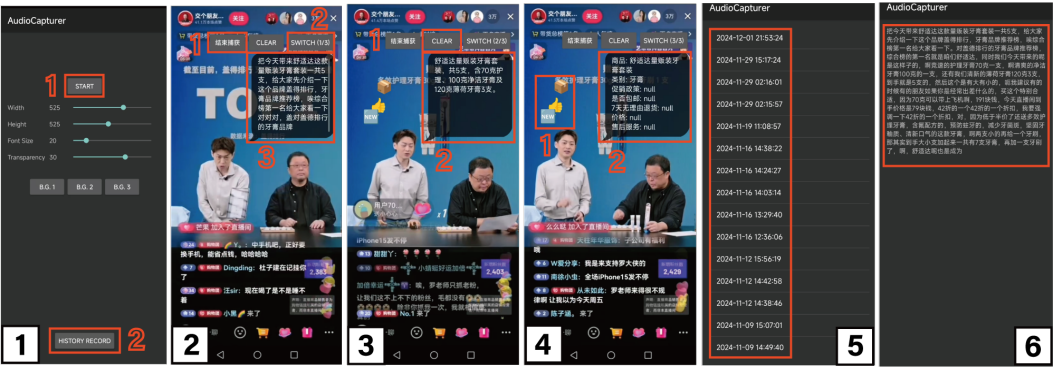

Fig. 4. *EchoAid* Floating Window Interface Functions: (1) Click to activate the software floating window mode. 1-2: Open History Saved Text Folder button, displays saved text history when clicked. (2) 2-1: Start/End Capture button, initiates/stops recording for in-camera sound translation subtitles. 2-2: Switch Interface button, toggles between interfaces (three available). 2-3: Raw Text Display interface (1/3); (3) 3-1: Clear button, refreshes all text contents for a new start and restarts recording. 3-2: Condensed Text Display interface (2/3) (4) 4-1: GUI interface, 4-2: Summary Framework Display interface (3/3); (5) Displays start times of each *EchoAid* usage section; click to show saved original text at that time. (6) Displays automatically saved original text information during use.

**Raw Text Display:** This view presents the raw speech-to-text output, enabling users to read the complete text of the livestream shopping episode (Figure7 (2)-(2-3)).

The iFlytek API ensures minimal delay for DHH users, enabling instant access to live voice information.

*4.2.2 Text Information Processing Phase.* After the initial speech-to-text conversion, *EchoAid* utilizes the *ERNIE Bot*, a large language model (LLM), to refine and distill the translated information every 30 seconds (D2). More detail please see Appendix A.2. *ERNIE Bot* processes the textual data in real-time, enhancing clarity and relevance of the presented information (D2). The refined results are then displayed on the *EchoAid* interface in three distinct formats (D4) to cater to the diverse needs and preferences of DHH users:

**Condensed Text Display:** Utilizing the RSVP method to display text word-by-word or character-by-character (Figure7 (3)-(3-2)), with refined and condensed content processed by *ERNIE Bot* (D3).

**Summary Framework Display:** This mode highlights key content using a structured summary framework tailored for livestream shopping (D2). Prompted by inputs to *ERNIE Bot* derived from extensive expert research and DHH user interviews, the system focuses on eight critical dimensions of livestream shopping information influencing user decisions (D4): promotional policies, free shipping availability, return policy, pricing, after-sales service, product introduction, usage experience, and user manual (Figure7 (4)-(4-2)). This selection aligns with the S-O-R model, recognizing these factors as key stimuli eliciting specific behavioral responses in the shopping process. The framework dynamically updates and accumulates content based on the host's narration, ensuring DHH users have access to the most relevant and crucial information in real time.

*4.2.3 User Interaction phase.* In the User Interaction Phase, *EchoAid* allows users to adjust the display interface features for a personalized experience. Users can freely move and resize the floating window to suit their preferences (D5). This customization enables users to position the window anywhere on the screen without blocking the live broadcast content, adapting to different usage scenarios and personal habits.

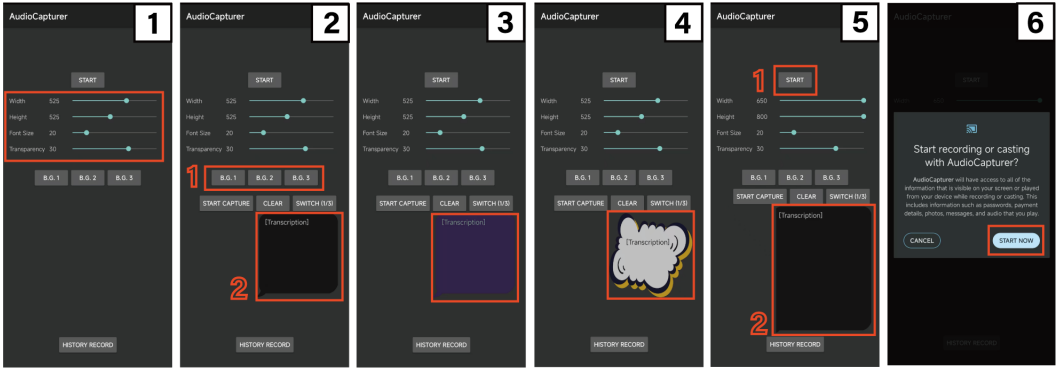

Fig. 5. Interface customization options in *EchoAid*: (1) Slide from top to bottom to adjust the width and height of the floating window, the size of subtitles, and the transparency of subtitles; (2, 3, 4) Customize the floating window with three different background choices. Button 2-1 selects the background for display 3, while 2-2 sets the background for display 1; (5) Once the floating window settings are configured, click on 5-1 to navigate from the interface to display 6; (6) Click the button to switch to display 5-2, activating the software floating window mode.

We developed a specialized GUI tailored for interactive enjoyment. This innovative design, depicted in Figure7 (4)-(4-1) (D6). The GUI dynamically displays icons corresponding to product details as the host discusses them, updating alongside the RSVP text box. Each icon represents different information categories like pricing, specifications, or user reviews, presenting data clearly and segmented. These intuitive GUI elements help *EchoAid* present complex information in a straightforward and engaging manner.

*4.2.4 Deployment.* *EchoAid* is developed using *Android* Studio and primarily supports *Android* 10.0 and higher versions. The app is developed in Kotlin to boost application security and efficiency. It comprises two primary interfaces: the Main Activity, responsible for permissions and user configurations, and the Floating Window Activity, which overlays text and video content and interacts with speech-to-text and LLMs. Furthermore, a foreground service is employed to handle audio processing and data transmission, ensuring smooth operation on modern *Android* devices.

## 4.3 Technical evaluation

We organized a user trial workshop and system reliability tests to assess the performance of the *EchoAid* prototype.

*4.3.1 User trial workshop.* We conducted a small-scale user trial workshop for *EchoAid* (version 1.0) with three DHH participants. The participants included a 32-year-old male vice president of a Deaf Association knowledgeable in accessibility technologies, a 25-year-old female school staff member active in livestream shopping, and a 24-year-old male digital product enthusiast keen on livestream shopping platforms. Each participant was given a *Huawei nova 5 Pro* smartphone running *HarmonyOS 4.0.0* with *EchoAid* software preinstalled for testing. They watched live broadcasts on platforms like *Douyin* for five minutes per session, totaling two sessions. Following the trial, we conducted semi-structured interviews with each user to explore the software's benefits and areas for improvement.

All users strongly endorsed design points D2, D3, and D4, stressing the significance of condensing content and extracting key information to reduce information overload. Users 2 and 3 expressed their appreciation for design aspects D1 and D5. User 2 highlighted the lack of customization options on other platforms, noting how *EchoAid*'s customizable features significantly improved her user experience. User 1 recommended enhancing the speed of keyword extraction and adding the ability to save translated texts locally. User 1 also provided suggestions on enhancing the entertainment value of livestream shopping platforms by proposing some GUI improvements, such as enabling users to customize the overlay window background.

Following the technical trial workshop, we identified areas for upgrading and iterating *EchoAid* (version 2.0) in the following dimensions:

**Dimension (1) User Interface Customization:** Incorporate additional customization options into the user interface. This includes enabling users to personalize the size and background of overlay windows.

**Dimension (2) Local Storage Function:** Introduce a new feature that allows users to store translated texts locally on their phones.

**Dimension (3) Enhanced Interactivity and Entertainment:** Enhance interactivity and entertainment by integrating more interactive features like gamification elements or interactive Q&A sessions to elevate the enjoyment level of watching live broadcasts.

*4.3.2 System reliability evaluation.* We gathered 20 livestream videos across four product categories (fruits, cosmetics, apparel, and furniture) from five prominent Livestream platforms in mainland China (*PDD*, *Kwai*, *Douyin* (Tiktok in mainland China), *Taobao*, and *Xigua Video*). Each livestream video was sampled ten times within five minutes (every 30 seconds). Subsequently, we assessed the performance of the *EchoAid* system on this sample. The test criteria and Examples please see appendix A.1.

For each sampling instance, 1 point was awarded if the information was accurately displayed on both interfaces 2 (Condensed Text Display) and 3 (Summary Framework Display); otherwise, 0 points were given. A perfect score of 10 indicates that throughout the ten sampling instances within the five minutes, all information on the Condensed Text Display and Summary Framework Display interfaces was correct for this livestream video.

Overall, *EchoAid* performed well in most tests, achieving a recognition accuracy of over 70% in most cases (15 out of 20), as depicted in Table 2. The primary errors observed were related to homophones in speech recognition and challenges in distinguishing between products and their subcategories, such as the similarity in the Chinese pronunciation of "9 pounds" and "Alcohol," leading to errors. These challenges were particularly prominent in the fruit category due to the diverse product names and lack of standardization, impacting recognition reliability. Despite these issues, *EchoAid* exhibited strong compatibility across various platforms, showcasing its generalizability and adaptability.

## 4.4 System iteration

Following feedback from DHH users in the technical trial workshop, we implemented two improvements to the *EchoAid* system:

**Customizable Comic Panel Backgrounds** Several DHH platform users (users 1, 2, 3, 4, 6, 7, 8) expressed that livestream shopping platforms are not attractive due to lack of sound reception. We introduced a comic-style background for the RSVP display of refined text (as depicted in Figure 5 (2)-(4)), presenting information in a more vivid and engaging manner.

Table 2. Text-to-speech reliability rating (RSVP and summary modes): Recognition test results of *EchoAid* across five live streaming platforms and four product categories. Each video is rated on a scale of 10, with a higher score indicating better recognition accuracy. Additionally, the table includes average scores across platforms and categories.

| Platform | Fruits | Make-ups | Clothes | Furniture | Avg |
|---|---|---|---|---|---|
| *PDD* | 7 | 8 | 9 | 6 | 7.50 |
| *Kwai* | 7 | 8 | 9 | 10 | 8.50 |
| *Douyin* | 5 | 9 | 3 | 10 | 6.75 |
| *Taobao* | 1 | 10 | 7 | 10 | 7.00 |
| *Xigua Video* | 4 | 10 | 10 | 10 | 8.50 |
| **Avg** | 4.80 | 9.00 | 7.60 | 9.20 | **7.65** |

**Text Local Save and Review Feature** Feedback from User 1 and User 2 emphasized the importance of saving text content locally for future reference. In response, we integrated an automatic text saving and review function into the *EchoAid* system, shown in Figures 7 (5), (6). Users can now save translated text directly to their phone's local storage and access it conveniently whenever needed.

## 5 STUDY III: *ECHOAID* USER STUDY

Following the iteration of the system, we plan to evaluate the performance of *EchoAid* in enhancing DHH users' livestream shopping experiences. The evaluation will specifically assess the effectiveness of *EchoAid* in reducing information overload during livestream shopping.

### 5.1 Participants

Same as in Study I, participants were recruited through online social media, provincial Disabled Persons' Federation, and NGOs. We recruited 38 DHH participants, with an average age of 30.26 (SD = 7.53) (refer to Table 3 for participants' demographic details). The recruitment process had three criteria: participants must be able to travel to the user research site, they must be DHH individuals, and they should have experience using livestream shopping platforms. Each DHH participant received a cash reward of 100 RMB as a token of appreciation for their time and contribution.

### 5.2 Apparatus

*EchoAid* is implemented on the *Huawei nova 5 Pro* utilizing *HarmonyOS 4.0.0* and is compatible with *Android* devices running versions 10 or higher. This setup enables real-time support for DHH users across different live commerce platforms. To maintain the experiment's integrity and ensure data reliability, local videos sourced from prominent livestream shopping platforms in China are employed. This method guarantees that all participants have access to identical content, eliminating discrepancies arising from network fluctuations or variations in streaming quality.

### 5.3 Setup and Experimental Design

The study involved 38 participants who were randomly divided into two groups. One group utilized *EchoAid*, while the other group used *iFlytek* to watch the same livestream shopping video playback, as illustrated in Figures 6 (4) and (5). Prior to commencing the tests, participants in the *EchoAid* group received a 5-10 minute tutorial to familiarize themselves with the system's functionalities.

Table 3. **Demographic Data of 38 DHH Participants. Age: 0~19, 20~30, 31~40, 41~50, 51~60. DHH State: Mild(speech audible with difficulty in subtle sounds), Moderate(loud or close-up speech needed), Moderately severe(dialogue inaudible, only very loud sounds heard), Severe(near-total deafness, requires aids or implants), Totally deaf(no hearing). Livestreaming Usage (How often do you usually watch content on the live streaming platform): Daily, Weekly, Monthly, Occasionally, Never.**

|  | Gender | Age | DHH State | Livestreaming Usage |  | Gender | Age | DHH State | Livestreaming Usage |
|---|---|---|---|---|---|---|---|---|---|
| A1 | F | 20~30 | Severe | Occasionally | B1 | F | 20~30 | Severe | Never |
| A2 | F | 20~30 | Moderately severe | Monthly | B2 | M | 20~30 | Severe | Never |
| A3 | F | 31~40 | Totally deaf | Weekly | B3 | F | 51~60 | Severe | Weekly |
| A4 | F | 31~40 | Severe | Weekly | B4 | M | 20~30 | Severe | Monthly |
| A5 | F | 31~40 | Severe | Monthly | B5 | F | 51~60 | Severe | Weekly |
| A6 | M | 41~50 | Moderately severe | Daily | B6 | F | 20~30 | Severe | Monthly |
| A7 | M | 51~60 | Moderately severe | Monthly | B7 | F | 20~30 | Severe | Never |
| A8 | M | 31~40 | Moderate | Weekly | B8 | F | 20~30 | Severe | Weekly |
| A9 | M | 31~40 | Severe | Weekly | B9 | F | 31~40 | Severe | Monthly |
| A10 | M | 20~30 | Moderate | Monthly | B10 | F | 31~40 | Moderately severe | Never |
| A11 | M | 0~19 | Severe | Weekly | B11 | F | 20~30 | Moderately severe | Monthly |
| A12 | F | 20~30 | Moderately severe | Monthly | B12 | F | 51~60 | Severe | Weekly |
| A13 | F | 31~40 | Moderately severe | Weekly | B13 | F | 20~30 | Severe | Never |
| A14 | F | 31~40 | Severe | Monthly | B14 | F | 20~30 | Severe | Weekly |
| A15 | M | 31~40 | Severe | Weekly | B15 | F | 51~60 | Severe | Monthly |
| A16 | F | 31~40 | Totally deaf | Never | B16 | M | 51~60 | Moderately severe | Monthly |
| A17 | F | 20~30 | Severe | Monthly | B17 | M | 31~40 | Severe | Monthly |
| A18 | F | 20~30 | Severe | Weekly | B18 | M | 20~30 | Moderately severe | Monthly |
| A19 | M | 20~30 | Severe | Monthly | B19 | M | 20~30 | Moderately severe | Weekly |

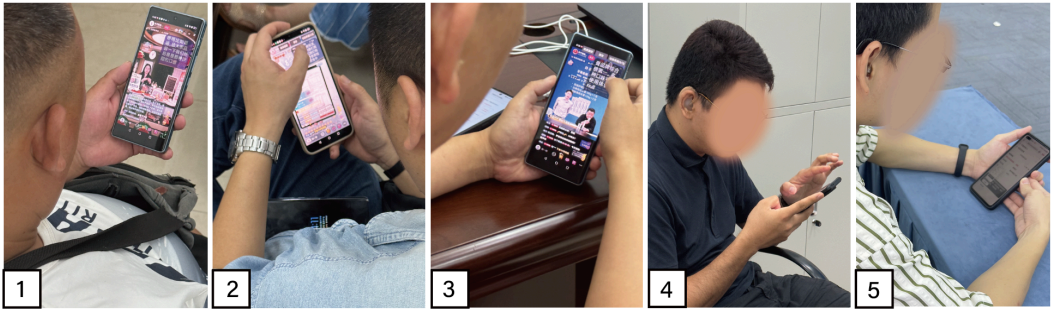

Fig. 6. Participants' operation with *EchoAid*: (1) Raw text display interface; (2) Condensed Text display interface; (3) Summary framework display interface; (4) Using *EchoAid* to assist in watching live e-commerce platforms; (5) Using *iFlytek* to assist in viewing live e-commerce platforms.

Figure 7 illustrates the operational flow of the two groups. In the *iFlytek* group (top figure), cognitive overload attention means that users are required to simultaneously focus on subtitles and visual information, leading to a high risk of information loss and information overload-related stress. In the *EchoAid* group (bottom figure), distributed attention means that users can allocate their attention more flexibly because of the condensed subtitle that is refreshed every 30 seconds. This allows more room for users to shift their focus between subtitles and visual content without feeling overwhelmed.

Participants needed to complete two tasks: (Task 1) Participants first watched a video selling watermelons (Video 1) and then completed a system experience questionnaire to assess the system's design and a NASA-TLX questionnaire to evaluate cognitive load.

**System Experience Questionnaire.** This questionnaire contains six dimensions– "usability," "ease of use," "content comprehensibility," "content memorability," "accuracy of summaries", and

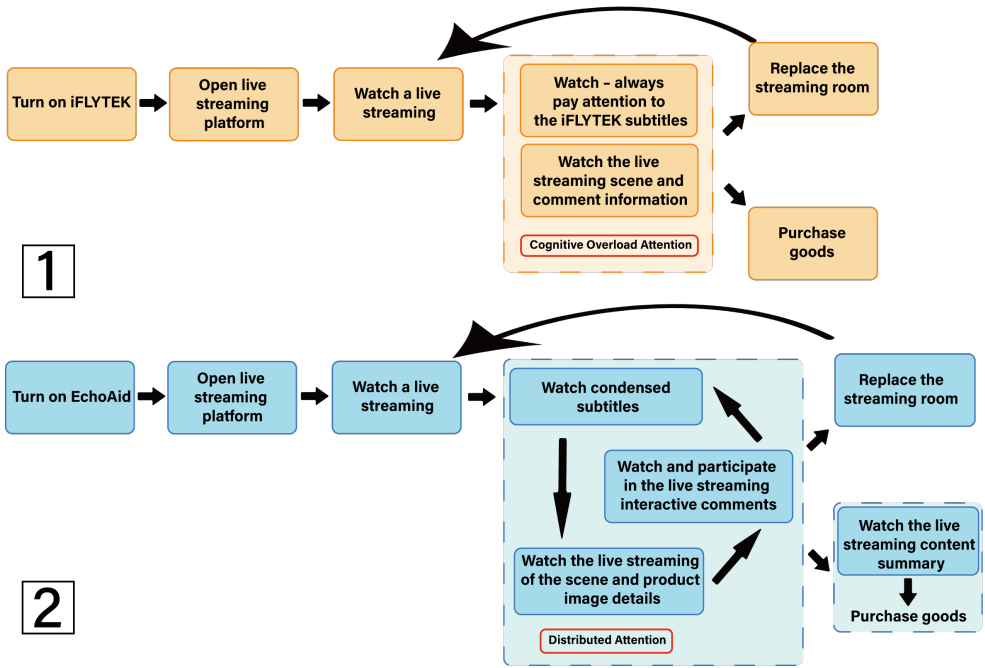

Fig. 7. Operational flow in the *iFLYTEK* group (top) and in the *EchoAid* group. (1) *iFLYTEK* subtitles constantly appear on the screen along with the host's speech, sometimes covering product images and host's facial expression and gestures. Users have to constantly check the subtitles when viewing the livestream. (2) *EchoAid* provides automatically condensing subtitle text every 30 seconds. Users could finish reading in seconds and allocate their time to other visual cues. Additionally, *EchoAid* provides a key information summary framework, allowing users to review the entire livestream content after it ends and make more informed purchase decisions.

"viewing interest" of the system, each dimension consisting of two questions. Users are asked to fill out the form after watching the first video. The questionnaire is attached in the Appendix Table (A1)

**NASA-TLX Questionnaire.** This questionnaire gathered subjective feedback on mental, physical, and temporal demands, performance, effort, and frustration, as reported in the Appendix Table (A2).

(Task 2) Subsequently, they watched two more videos, one selling latex mattresses (Video 2) and another selling toothpaste (Video 3). After viewing all two videos, a memory test was conducted to evaluate information retention. Both groups, one using *EchoAid* and the other *iFlytek*, underwent the same condition. The video content and test questions were developed with input from two experts: an assistant professor specializing in accessibility research and an industry expert in *Douyin* live commerce.

**Livestream shopping Memory Test.** The test focused on product features, discount policies, price changes, return policies, and product functionality introduction. The questionnaire had 10 questions, with 2 questions per dimension and 4 answer options each. Participants earned 1 point for each correct answer, with no points deducted for incorrect or unanswered questions. To pass, participants needed to score 5 points or more. The questionnaire is attached in the Appendix Table (A3)

**Interviews.** After the questionnaires were completed, we randomly chose ten participants for semi-structured interviews, with five from the baseline group and five from the *EchoAid* group. Through the content analysis of these interviews, we identified key themes to guide the future enhancement of *EchoAid*.

### 5.4 Data Analysis

We used the Mann-Whitney U (MWU) test [51] to compare test score differences between the *iFlytek* (baseline) and *EchoAid* groups. The results from the System Experience Questionnaire, NASA-TLX Questionnaire, and Live Information Memory Test are presented in Figure 8, Figure 9, and Figure 10, respectively.

## 6 RESULTS

### 6.1 System Experience Questionnaire Results

Figure 8 illustrates the outcomes across six dimensions (corresponding to twelve questions) under two conditions. We report the detailed results of the questionnaire below.

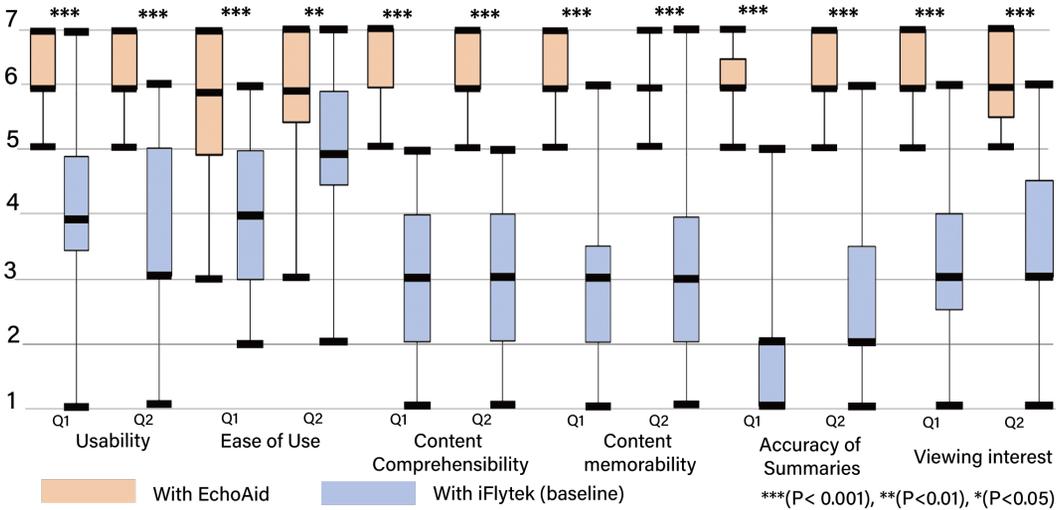

Fig. 8. Participants' ratings on the system experience questionnaire: X-axis: Participants' ratings on Usability, Ease of Use, Usability, Content comprehensibility, Memorability of content, Accuracy of summaries, and Viewing interest; Y-axis: Ranging from 1 (Strongly Disagree) to 7 (Strongly Agree).

For **Usability**, Q1 "*This system is very helpful to me when watching livestream shopping*": Compared to baseline *iFlytek* ($\bar{M}$ = 4.26, SD = 1.55), participants using *EchoAid* felt the system provided better help ($\bar{M}$ = 6.42, SD = 0.61). The result of the MWU test is ($U$ = 327.5, p <0.001). Q2 "*Using this system, my livestream shopping experience has become easier*": The livestream shopping experience of *EchoAid* ($\bar{M}$ = 6.32, SD = 0.67) is better than that of the *iFlytek* group ($\bar{M}$ = 3.68, SD = 1.41). The MWU test result is ($U$ = 343.0, p <0.001).

For **Ease of Use**, Q1 "*I find this system's interface user-friendly and the functions easy to use.*" indicates the participants using *EchoAid* feel it is easier to use ($\bar{M}$ = 5.79, SD = 1.18) compared *iFlytek* ($\bar{M}$ = 4, SD = 1.24). The result of the MWU test is ($U$ = 304.5, p <0.001). Q2 "*I can quickly* *get started and operate this system to watch livestream shopping*" reflects an obvious difference ($U$ = 267.5, p <0.01). Compared to the *iFlytek* group ($\bar{M}$ = 4.84, SD = 1.42), participants using *EchoAid* found it easier to navigate during livestream shopping ($\bar{M}$ = 5.95, SD = 1.13).

For **Content Comprehensibility**, Q1 "*Using this system makes it easier for me to understand the livestream content.*" : Compared to participants using *iFlytek* ($\bar{M}$=3.16, SD=1.26), those using *EchoAid* found it easier to understand content during livestream shopping ($\bar{M}$ = 6.32, SD = 0.82). The result of the MWU test is ($U$ = 353.0, p <0.001). Q2 "*I can better keep up with the pace of the facilitator through this system.*": Participants using *EchoAid* were better able to follow the facilitator's tempo ($\bar{M}$ = 6.32, SD = 0.58) compared to participants using *iFlytek* ($\bar{M}$ = 3.05, SD = 1.08). The MWU test result is ($U$ = 360.0, p <0.001).

Regarding **Content memorability**, Q1 "*This system allows me to more easily remember important livestream information.*" : Participants using *iFlytek* ($\bar{M}$ = 3.05, SD = 1.18) and *EchoAid* ($\bar{M}$ = 6.37, SD = 0.68) had differing levels. The MWU test result is ($U$ = 354.0, p <0.001). Q2 "*Using this system, I can clearly recall what the facilitator discussed.*" : Compared to participants using *iFlytek* ($\bar{M}$ = 3.16, SD = 1.46), those using *EchoAid* felt they could better understand the content by the facilitator ($\bar{M}$ = 6.00, SD = 0.67). The MWU test result is ($U$ = 340.0, p <0.001).

Next, **Accuracy of Summaries**, Q1 "*This system is very accurate in summarizing the content of livestream shopping.*" compared to *iFlytek* ($\bar{M}$ = 2.05, SD = 1.08), participants using *EchoAid* ($\bar{M}$ = 6.16, SD = 0.60) were able to better understand the content summarized during livestream shopping, with a significant difference ($U$ = 360.0, p <0.001). Q2 "*This system can help me quickly grasp the core information during the livestream shopping.*" : *EchoAid* participants ($\bar{M}$ = 6.21, SD = 0.63) scored higher than *iFlytek* participants ($\bar{M}$ = 2.84, SD = 1.38). The MSU test result is ($U$ = 351.5, p <0.001).

Finally, regarding the **Viewing interest**, Q11 "*This system makes me feel more interested in watching livestream shopping.*", compared to the baseline *iFlytek* ($\bar{M}$ = 3.26, SD = 1.24), participants using *EchoAid* felt more interested in livestream shopping ($\bar{M}$ = 6.16, SD = 0.69). The MWU test result is ($U$ = 350.0, p <0.001). Q12 "*After using this system, I will be more inclined to watch livestream shopping.*". Compared to participants using *iFlytek* ($\bar{M}$ = 3.42, SD = 1.30), *EchoAid* participants are more likely to conduct livestream shopping in the future ($\bar{M}$ = 6.11, SD = 0.81). The MWU test result is ($U$ = 342.5, p <0.001).

## 6.2  NASA-TLX Questionnaire Result

Figure 9 illustrates the results for the six dimensions of the NASA-TLX under two conditions: using *EchoAid* and using *iFlytek*. Across all dimensions, the scores for participants using *EchoAid* were significantly lower, indicating reduced cognitive load compared to *iFlytek*. Notable differences were observed in Mental Demand ($U$ = 99.0, p <0.05), Temporal Demand ($U$ = 74, p <0.01), Effort ($U$ = 66.0, p <0.001), and Frustration ($U$ = 76.5, p <0.01).

For Mental Demand, *EchoAid* users reported lower scores ($\bar{M}$ = 38.89, SD = 23.31) compared to *iFlytek* users ($\bar{M}$ = 63.11, SD = 17.65), demonstrating a statistically significant reduction in perceived mental effort. Similarly, Temporal Demand was significantly lower with *EchoAid* ($\bar{M}$ = 35.84, SD = 28.42) than *iFlytek* ($\bar{M}$ = 66.16, SD = 18.42), reflecting an improved ability to process and follow real-time information effectively.

In addition, for Effort, participants using *EchoAid* reported significantly lower scores ($\bar{M}$ = 35.79, SD = 24.86) compared to the baseline *iFlytek* ($\bar{M}$ = 66.47, SD = 22.18), indicating a clear reduction in the effort required. Meanwhile, in the Frustration dimension, *EchoAid* users achieved substantially lower scores ($\bar{M}$ = 23.42, SD = 26.40) than *iFlytek* users ($\bar{M}$ = 51.42, SD = 26.43), further emphasizing *EchoAid*'s effectiveness in alleviating user frustration during the livestream shopping experience.

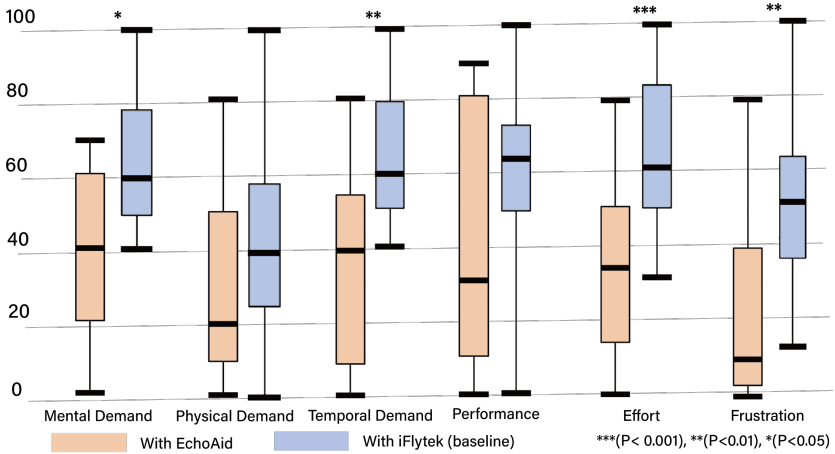

Fig. 9. Participants' ratings on the NASA TLX scale. X-axis: Six dimensions assessing the psychological load of the task, including Mental Demand, Physical Demand, Temporal Demand, Performance, Effort, and Frustration; Y-axis: Ranging from 0 to 100, lower scores indicate better performance.

### 6.3 Live Information Memory Test

Among the 19 participants using *EchoAid*, 17 scored 5 or higher (out of 10) ($\bar{M}$=6.53, SD=1.71), as shown in Figure 10. 11 participants using *iFlytek* achieved scores above 5 ($\bar{M}$=5.32, SD=1.97). As shown in Figure 10, *EchoAid* marginally outperforms the baseline method in terms of memory retention for live stream content. Participants using *EchoAid* tended to remember more details about products presented by hosts.

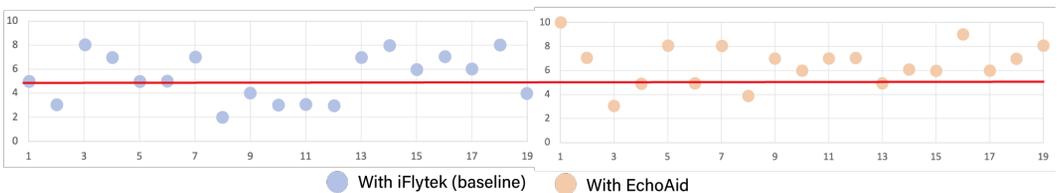

Fig. 10. Test scores for Live information memory. X-axis: Participant IDs, and Y-axis: score. The red line in the middle indicates the passing threshold—only scores of 5 or higher are considered passing.

### 6.4 DHH Participants Feedback

We interviewed 10 participants randomly selected from the 19 participants in the *EchoAid* user experiment (numbered as a1, a2, a5, a7, a9, a11, a13, a15, a16, a18). Participants unanimously praised *EchoAid*'s automatic keyword extraction and condensed information display functions, considering them the most valuable features. Users a1, a2, a5, a7, and a9 mentioned that these functions enabled them to quickly locate and understand critical information, allowing them to better keep up with the streamer. For example, user a2 stated, *"With EchoAid, I no longer worry about missing important information from the host. The software automatically highlights keywords, helping me stay focused even when multitasking."* User a11 added, *"This tool helps me quickly focus on the key information without having to sift through lengthy subtitles."*

Users a11, a13, a15, a16, and a18 also provided positive feedback on the system's real-time transcription and interface design. User a13 noted, *"The ability to switch between subtitle modes is very convenient and allows for choosing either condensed or original mode depending on the scenario."* User a16 mentioned, *"The real-time speech-to-text and automatic keyword extraction features are very practical, especially in fast-paced livestreams, saving a lot of time."* User a18 remarked, *"The condensed subtitle mode does an excellent job of summarizing, reducing cognitive load, and allowing me to keep up with the livestream content effortlessly."*

Participants also proposed suggestions for improving *EchoAid*'s functionality to further optimize its performance. User a9 suggested, *"1. Improve the language recognition model to avoid mixing English and Chinese; 2. Add bilingual recognition; 3. Enhance the user-friendliness and accessibility of function setting icons."* User a5 recommended adding livestream playback and automatic subtitle-saving features to facilitate reviewing key information. User a13 proposed, *"If subtitles could be linked directly to product pages, allowing users to click on keywords to quickly access product details, it would make the shopping experience more efficient."* Users a7 and a15 further stated: *"If EchoAid could incorporate a sign language-to-speech feature, it would provide a richer livestream interaction experience for hard hearing users,"* and *"allow hearing audiences to easily understand the content made by sign language livestreamer, significantly expanding EchoAid's application scope and value."* More detail see in Appendix A.4.

It is evident that *EchoAid* significantly reduces information loss and overload while enhancing the overall experience of livestream shopping. This improvement is largely attributed to *EchoAid*'s real-time, context-aware keyword extraction and information display capabilities, which effectively simplify information processing tasks for users. Participants consistently expressed that these features were instrumental in helping them navigate the fast-paced, information-dense livestream shopping environment. In the future, *EchoAid* will focus on implementing these user suggestions, continuously optimizing system performance, and expanding its functionalities to better serve the needs of the hearing-impaired community.

## 7  DISCUSSION

We conducted three studies to support and validate the development of *EchoAid*, aiming to enhance the livestream shopping experience for DHH users. Guided by the S-O-R model, we systematically formulated our interview questions to uncover the specific challenges faced by DHH users in livestream shopping, with a particular focus on information loss and overload. By structuring our inquiry around the stimuli (e.g., hosts' speech, rapid promotional updates, and interactive visuals), the organism (DHH users' cognitive processing and accessibility barriers), and the response (their engagement and decision-making behaviors), we were able to pinpoint key pain points that impact their shopping experience. This model informed the prime goal of our *EchoAid* to reduce the cognitive overload via minimizing information loss and information overload. Specifically, we integrated real-time speech-to-text conversion, keyword extraction via LLMs, and RSVP-based display strategies to streamline information presentation and minimize cognitive overload.

*EchoAid* effectively addresses information acquisition barriers on live commerce platforms for DHH users by employing real-time, context-aware keyword extraction and summary information display. This enhances the accessibility and user interaction experience of assistive software on live commerce platforms. In this section, we also explore the potential extension of *EchoAid* in different user interfaces, its compatibility with various live commerce platforms, and its adaptation to a wider range of public touchscreen devices. Lastly, we situate this study within the HCI and CSCW literature, review current limitations and propose potential future research directions.

## 7.1 Structured Sale Information Extraction to Reduce Information Loss and Overload

To our knowledge, *EchoAid* is the first application tailored specifically to aid the DHH community in livestream shopping. Unlike conventional tools like *iFlytek* that focus solely on basic speech-to-text conversion, *EchoAid* stands out by not only converting speech to text but also presenting the sale information in a structured format. This highly structured sale informative presentation approach effectively helps DHH users capture sale information, mitigating issues related to information loss and information overload. The concise text display elevates the livestream shopping experience for DHH users, improving the accessibility and interactivity of the user interface.

## 7.2 Episode-Specific Attention to Enhance Comprehension

*EchoAid* incorporates a carefully designed error recovery support system. A standout feature of *EchoAid* is the one-click clear screen button, which enables users to swiftly clear all text content from the screen when transitioning between products or livestream episodes. This functionality prompts the Large Language Models (LLMs) to restart text processing, ensuring that the analysis begins anew with fresh livestream content, preventing misunderstanding due to interference from previous information. This operational principle is based on the real-time speech-to-text conversion of livestream audio into text, which is then inputted into the LLMs for information condensation and keyword extraction [17].

In our system implementation during Study II, we observed that when the host switches product sale or when the viewer changes livestream episodes, the accumulated text may contain disjointed or irrelevant information. This switch can result in errors in LLM processing, such as extracting incorrect keywords. By utilizing the one-click clear screen feature, users can proactively eliminate confusing inputs when needed, resetting the information processing flow to allow the LLMs to analyze only the most recent and relevant content. This error recovery mechanism not only enhances the system's usability and reliability but also improves the overall user experience. Moving forward, we aim to enhance this feature further by considering the automatic triggering of this refreshing function based on the smart detection of product changes during promotions.

## 7.3 Customized Interface to Support Individual Needs and Preferences

*EchoAid* provides DHH users with highly customized interface options, reflecting a deep understanding and consideration of the needs of the DHH community. Our system offers a range of customizable display settings to accommodate individual preferences, including Raw Text Display, Condensed Text Display, and Summary Framework Display. Users can adjust subtitle font sizes, floating window sizes, positions, and backgrounds to optimize visual clarity and reading experiences. Additionally, *EchoAid* allows users to personalize the graphical user interface (GUI) and floating window backgrounds, enhancing information accessibility and visual presentation customization. This level of customization empowers DHH users to tailor how information is presented based on their viewing habits while ensuring they do not miss any critical information.

The significance of personalization and user customization has been emphasized in prior research studies [3, 13]. Moving forward, we aim to explore additional personalization and interactive features to enrich the user experience and increase user satisfaction. For instance, we are introducing new interactive elements such as customizable icons, emojis, and buttons that can be positioned and resized on the interface and linked to specific functions, like accessing common functions swiftly or adjusting text flow speed and density. This interactive approach aims to enhance user control and engagement in livestream shopping, enabling DHH users to actively participate in streams rather than passively consume information.

## 7.4 Extension to other scenarios

While *EchoAid* has demonstrated significant benefits for the DHH community in livestream shopping scenarios, its potential applications extend far beyond this context. The core design of *EchoAid*, which integrates LLMs and speech-to-text functionality for information processing and summarization, lends itself well to various entertainment platforms catering to DHH users, including regular livestreams, video playback, movie watching, online meetings, and gaming interactions [6, 9, 29]. Furthermore, *EchoAid* can be adapted for everyday communication, serving as a valuable tool for DHH individuals. Specifically, in educational settings or work meetings where DHH individuals need to juggle visual information from speakers and subtitles, *EchoAid* can assist by transcribing lectures and discussions in real-time, extracting key points, and facilitating more effective information acquisition for DHH students.

Moving forward, our plan is to further develop and optimize *EchoAid*'s functionalities to meet specific needs and address complex environments. This includes adding support for multiple languages, enhancing recognition accuracy in noisy environments, and improving directional voice recognition capabilities. Through these technological advancements and expanded applications, *EchoAid* is poised to play a vital role in helping DHH users overcome communication barriers and enhance their overall quality of life.

## 7.5 *EchoAid* in Conjunction with Other DHH Assistive Technologies

We are currently exploring the integration of *EchoAid* with various assistive technologies, including hearing aids [63], visual aids [41], and cutting-edge technologies such as Virtual Reality (VR) [38, 53, 80, 85], Augmented Reality (AR) [23, 49], and Mixed Reality (MR) [78, 86].

Through the use of VR, AR, and MR [37, 76, 79, 84], we can create personalized entertainment experiences tailored to DHH users. For instance, in cinemas, AR technology could project real-time subtitles directly into the field of view of DHH individuals [5], while MR technology could translate significant sound effects from movies into visual or tactile signals [32]. This innovative approach allows DHH users to engage with movies in new and exciting ways. Integrating *EchoAid* with other assistive technologies empowers DHH users to choose the speech-to-text/condensed text or other visual/tactile cues they prefer, enhancing their confidence and convenience and unlocking new possibilities for the DHH community. The combination of these technologies aims to provide a more comprehensive and immersive experience for the DHH community.

## 7.6 Limitation and Future work

One general limitation of our study is that all participants were recruited online and had prior experience with platform usage, which might not represent the broader DHH community. These DHH platform users may be more open towards the use of livestream shopping platforms compared to non-users and are more advanced in accepting new technologies. Future research should involve a wider range of DHH users, including non-users of livestream shopping platforms, to ensure more comprehensive insights. Furthermore, the speech-to-text technology coupled with LLMs did not effectively capture the nuances in tone and emotion embedded in the livestream host's language, potentially diminishing engagement with the host and other viewers. Therefore, exploring alternative models that consider the tone of voice, such as multimodal LLMs that incorporate both speech and facial expressions, could enhance the quality of user engagement and interaction. In our user study, we did not assess the participants' purchasing decisions because the live stream scenarios were predetermined by the researchers, rather than reflecting the participants' actual shopping desires. Future studies should conduct more natural experiments, especially considering that the S-O-R model extends to eventual purchasing decisions. This approach will help us better understand the benefits of different systems in real-life settings.

**System robustness** *EchoAid* occasionally faces challenges in handling high information flow and complex visual environments, especially in noisy backgrounds or live settings with multiple hosts or multiple products[56]. In these scenarios, the system sometimes fails to extract key information accurately, which may result in missing important information. Considering these issues, future improvements will include integrating more effective LLMs and speech-to-text systems to enhance the capability of targeted information capture in complex environments [59].

Furthermore, our study primarily focuses on the Stimulus (S) and Organism (O) components of the S-O-R model, examining how livestream shopping stimuli impact the cognitive processing of DHH users. However, the Response (R) component, which reflects whether these cognitive effects translate into actual purchasing behavior, remains unexplored. Future research should investigate how improved accessibility interventions influence user engagement, decision-making, and ultimately, conversion rates in livestream shopping. Understanding this behavioral response would provide deeper insights into the effectiveness of assistive technologies in driving purchasing behavior among DHH consumers.

**Visual Feedback Design** When users quickly switch between livesteam episodes, they need to manually clear the recognized text and restart the recording, which can affect the system's response speed and accuracy. These issues partly stem from the limitations of the current speech recognition and keyword extraction algorithms used in the LLMs. Future developments could include the adoption of hybrid machine learning models to optimize algorithm performance [31], adapting to the dynamic nature of livestream viewing behaviors [43].

**Dialect** *EchoAid* currently struggles with handling non-standardized speech and dialect-rich expressions, which is particularly evident in livestream from certain regions. To address this, future efforts will focus on improving the language processing module to better understand dialects and slang [39, 73], thereby enhancing overall recognition accuracy. Currently, *EchoAid* is recommended for use with Chinese-language live commerce platforms, and it only supports Chinese subtitles.

**Interaction Challenges for Users with Different Levels of Hearing Impairment** During our user testing, we encountered a completely deaf user who primarily relied on sign language for communication. She is relatively older and has been deaf since birth. Having long relied on sign language for communication, she finds it challenging to fully comprehend the textual explanations provided by *EchoAid*. Instead, she prefers using images and visual aids to understand her surroundings. Therefore, to improve the experience for completely deaf users, *EchoAid* plans to integrate more visual elements and sign language translation features. This will include support for Avatar sign language captions or assistive features that convert sign language into text-based captions [8, 22]. These enhancements aim to create a more accessible and inclusive environment, making it easier for completely deaf users to engage with live content and interact more naturally within the platform.

## 8 CONCLUSIONS

Livestream shopping is becoming increasingly popular in China. Although livestream shopping platforms offer unprecedented interactivity and convenience, DHH users still face significant challenges in using them. Our development of the *EchoAid* tool is a valuable supplement to existing technology, combining advanced speech recognition technologies, LLMs, and RSVP technology to provide a more seamless and inclusive livestream shopping experience for the DHH community. Our research has shown that the application of *EchoAid* significantly improves the shopping experience for DHH users. Users can receive and process information more accurately with reduced cognitive load, and more effectively participate in livestream shopping interactions. Additionally, *EchoAid* demonstrates how technological innovation can address social inclusivity issues, especially in the rapidly developing field of e-commerce. Through continuous technological iteration and user feedback, we hope to make *EchoAid* the practical tool for supporting the DHH community in livestream shopping.

## 9 ACKNOWLEDGEMENT

This work is supported by the Guangdong Provincial Talent Program, Grant No.2023JC10X009.

## REFERENCES

[1] Wan Anita Wan Abas, Moniza Waheed, et al. 2024. The Influence Of Anchor Characteristics On Purchasing Behavior: A Systematic Review. *Migration Letters* 21, S2 (2024), 1038–1063.

[2] Khalid Nasser Alasim. 2019. Reading development of students who are deaf and hard of hearing in inclusive education classrooms. *Education Sciences* 9, 3 (2019), 201.

[3] Saifeddin Alimamy and Juergen Gnoth. 2022. I want it my way! The effect of perceptions of personalization through augmented reality and online shopping on customer intentions to co-create value. *Computers in Human Behavior* 128 (2022), 107105.

[4] Miriam Arnold, Mascha Goldschmitt, and Thomas Rigotti. 2023. Dealing with information overload: a comprehensive review. *Frontiers in Psychology* 14 (2023), 1122200.

[5] Grigoris Bastas, Maximos Kaliakatsos-Papakostas, Georgios Paraskevopoulos, Pantelis Kaplanoglou, Konstantinos Christantonis, Charalampos Tsioustas, Dimitris Mastrogiannopoulos, Depy Panga, Evita Fotinea, Athanasios Katsamanis, et al. 2022. Towards a DHH accessible theater: real-time synchronization of subtitles and sign language videos with ASR and NLP solutions. In *Proceedings of the 15th International Conference on PErvasive Technologies Related to Assistive Environments.* 653–661.

[6] Johnna Blair and Saeed Abdullah. 2019. Understanding the needs and challenges of using conversational agents for deaf older adults. In *Companion Publication of the 2019 Conference on Computer Supported Cooperative Work and Social Computing.* 161–165.

[7] Oleg Borisov, Mohammad Aliannejadi, and Fabio Crestani. 2021. Keyword extraction for improved document retrieval in conversational search. *arXiv preprint arXiv:2109.05979* (2021).

[8] Lauren Buck, Gareth W Young, and Rachel McDonnell. 2023. Avatar Customization, Personality, and the Perception of Work Group Inclusion in Immersive Virtual Reality. In *Companion Publication of the 2023 Conference on Computer Supported Cooperative Work and Social Computing.* 27–32.

[9] Janine Butler. 2019. Perspectives of deaf and hard of hearing viewers of captions. *American Annals of the Deaf* 163, 5 (2019), 534–553.

[10] Beiyan Cao, Changyang He, Muzhi Zhou, and Mingming Fan. 2023. Sparkling silence: Practices and challenges of livestreaming among deaf or hard of hearing streamers. In *Proceedings of the 2023 CHI Conference on Human Factors in Computing Systems.* 1–15.

[11] Jiaxun Cao, Xuening Peng, Fan Liang, and Xin Tong. 2024. "Voices Help Correlate Signs and Words": Analyzing Deaf and Hard-of-Hearing (DHH) TikTokers' Content, Practices, and Pitfalls. In *Proceedings of the CHI Conference on Human Factors in Computing Systems.* 1–18.

[12] Monica S Castelhano and Paul Muter. 2001. Optimizing the reading of electronic text using rapid serial visual presentation. *Behaviour & Information Technology* 20, 4 (2001), 237–247.

[13] Shobhana Chandra, Sanjeev Verma, Weng Marc Lim, Satish Kumar, and Naveen Donthu. 2022. Personalization in personalized marketing: Trends and ways forward. *Psychology & Marketing* 39, 8 (2022), 1529–1562.

[14] Chia-Chen Chen and Jun-You Yao. 2018. What drives impulse buying behaviors in a mobile auction? The perspective of the Stimulus-Organism-Response model. *Telematics and Informatics* 35, 5 (2018), 1249–1262.

[15] Si Chen, Haocong Cheng, Jason Situ, Desirée Kirst, Suzy Su, Saumya Malhotra, Lawrence Angrave, Qi Wang, and Yun Huang. 2024. Towards Inclusive Video Commenting: Introducing Signmaku for the Deaf and Hard-of-Hearing. In *Proceedings of the CHI Conference on Human Factors in Computing Systems.* 1–18.

[16] Yu-Chen Chen, Rong-An Shang, and Chen-Yu Kao. 2009. The effects of information overload on consumers' subjective state towards buying decision in the internet shopping environment. *Electronic Commerce Research and Applications* 8, 1 (2009), 48–58.

[17] Robert Chew, John Bollenbacher, Michael Wenger, Jessica Speer, and Annice Kim. 2023. LLM-assisted content analysis: Using large language models to support deductive coding. *arXiv preprint arXiv:2306.14924* (2023).

[18] Oscar De Bruijn, Robert Spence, and Min Yih Chong. 2002. RSVP browser: Web browsing on small screen devices. *Personal and Ubiquitous Computing* 6 (2002), 245–252.

[19] Martin J Eppler and Jeanne Mengis. 2008. The Concept of Information Overload-A Review of Literature from Organization Science, Accounting, Marketing, MIS, and Related Disciplines (2004) The Information Society: An International Journal, 20 (5), 2004, pp. 1–20. *Kommunikationsmanagement im Wandel: Beiträge aus 10 Jahren= mcminstitute* (2008), 271–305.

[20] Alet C Erasmus, Elizabeth Boshoff, and Gabriel G Rousseau. 2001. Consumer decision-making models within the discipline of consumer science: a critical approach. *Journal of Consumer Sciences* 29 (2001).

[21] Sevgin A Eroglu, Karen A Machleit, and Lenita M Davis. 2001. Atmospheric qualities of online retailing: A conceptual model and implications. *Journal of Business research* 54, 2 (2001), 177–184.

[22] Uzma Farooq, Mohd Shafry Mohd Rahim, Nabeel Sabir, Amir Hussain, and Adnan Abid. 2021. Advances in machine translation for sign language: approaches, limitations, and challenges. *Neural Computing and Applications* 33, 21 (2021), 14357–14399.

[23] Natália Fernandes, Antonio José Melo Leite Junior, Edgar Marçal, and Windson Viana. 2023. Augmented reality in education for people who are deaf or hard of hearing: a systematic literature review. *Universal Access in the Information Society* (2023), 1–20.

[24] Leah Findlater, Bonnie Chinh, Dhruv Jain, Jon Froehlich, Raja Kushalnagar, and Angela Carey Lin. 2019. Deaf and hard-of-hearing individuals' preferences for wearable and mobile sound awareness technologies. In *Proceedings of the 2019 CHI Conference on Human Factors in Computing Systems*. 1–13.

[25] Kenneth I Forster. 1970. Visual perception of rapidly presented word sequences of varying complexity. *Perception & psychophysics* 8 (1970), 215–221.

[26] Luther C Gilbert. 1959. Speed of processing visual stimuli and its relation to reading. *Journal of Educational Psychology* 50, 1 (1959), 8.

[27] Abraham Glasser, Joseline Garcia, Chang Hwang, Christian Vogler, and Raja Kushalnagar. 2021. Effect of caption width on the TV user experience by deaf and hard of hearing viewers. In *Proceedings of the 18th International Web for All Conference*. 1–5.

[28] Abraham Glasser, Vaishnavi Mande, and Matt Huenerfauth. 2020. Accessibility for deaf and hard of hearing users: Sign language conversational user interfaces. In *Proceedings of the 2nd Conference on Conversational User Interfaces*. 1–3.

[29] Cole Gleason, Patrick Carrington, Lydia B Chilton, Benjamin M Gorman, Hernisa Kacorri, Andrés Monroy-Hernández, Meredith Ringel Morris, Garreth W Tigwell, and Shaomei Wu. 2019. Addressing the accessibility of social media. In *Companion Publication of the 2019 Conference on Computer Supported Cooperative Work and Social Computing*. 474–479.

[30] Jan Gugenheimer, Katrin Plaumann, Florian Schaub, Patrizia Di Campli San Vito, Saskia Duck, Melanie Rabus, and Enrico Rukzio. 2017. The impact of assistive technology on communication quality between deaf and hearing individuals. In *Proceedings of the 2017 ACM Conference on Computer Supported Cooperative Work and Social Computing*. 669–682.

[31] Amin Ul Haq, Jian Ping Li, Muhammad Hammad Memon, Shah Nazir, and Ruinan Sun. 2018. A hybrid intelligent system framework for the prediction of heart disease using machine learning algorithms. *Mobile information systems* 2018, 1 (2018), 3860146.

[32] Tatsuya Honda, Tetsuaki Baba, and Makoto Okamoto. 2022. Ontenna: Design and social implementation of auditory information transmission devices using tactile and visual senses. In *International Conference on Computers Helping People with Special Needs*. Springer, 130–138.

[33] Md Alamgir Hossain, Abul Kalam, Md Nuruzzaman, and Minho Kim. 2023. The power of live-streaming in consumers' purchasing decision. *SAGE Open* 13, 4 (2023), 21582440231197903.

[34] Yue Huang and Lu Suo. 2021. Factors affecting Chinese consumers' impulse buying decision of live streaming E-commerce. *Asian Social Science* 17, 5 (2021), 16.

[35] Matt Huenerfauth and Vicki Hanson. 2009. Sign language in the interface: access for deaf signers. *Universal Access Handbook. NJ: Erlbaum* 38 (2009), 14.

[36] Andri Ioannou and Vaso Constantinou. 2018. Augmented reality supporting deaf students in mainstream schools: Two case studies of practical utility of the technology. In *Interactive Mobile Communication Technologies and Learning: Proceedings of the 11th IMCL Conference*. Springer, 387–396.

[37] Dhruv Jain, Leah Findlater, Jamie Gilkeson, Benjamin Holland, Ramani Duraiswami, Dmitry Zotkin, Christian Vogler, and Jon E Froehlich. 2015. Head-mounted display visualizations to support sound awareness for the deaf and hard of hearing. In *Proceedings of the 33rd Annual ACM Conference on Human Factors in Computing Systems*. 241–250.

[38] Dhruv Jain, Sasa Junuzovic, Eyal Ofek, Mike Sinclair, John R. Porter, Chris Yoon, Swetha Machanavajhala, and Meredith Ringel Morris. 2021. Towards sound accessibility in virtual reality. In *Proceedings of the 2021 International Conference on Multimodal Interaction*. 80–91.

[39] Bing-Hwang Juang and Sadaoki Furui. 2000. Automatic recognition and understanding of spoken language-a first step toward natural human-machine communication. *Proc. IEEE* 88, 8 (2000), 1142–1165.

[40] Saba Kawas, George Karalis, Tzu Wen, and Richard E Ladner. 2016. Improving real-time captioning experiences for deaf and hard of hearing students. In *Proceedings of the 18th International ACM SIGACCESS Conference on Computers and Accessibility*. 15–23.

[41] Muiz Ahmed Khan, Pias Paul, Mahmudur Rashid, Mainul Hossain, and Md Atiqur Rahman Ahad. 2020. An AI-based visual aid with integrated reading assistant for the completely blind. *IEEE Transactions on Human-Machine Systems* 50, 6 (2020), 507–517.

[42] Yeon Soo Kim, Sunok Lee, and Sangsu Lee. 2022. A Participatory Design Approach to Explore Design Directions for Enhancing Videoconferencing Experience for Non-signing Deaf and Hard of Hearing Users. In *Proceedings of the 24th International ACM SIGACCESS Conference on Computers and Accessibility*. 1–4.

[43] Jonathan Kua, Grenville Armitage, and Philip Branch. 2017. A survey of rate adaptation techniques for dynamic adaptive streaming over HTTP. *IEEE Communications Surveys & Tutorials* 19, 3 (2017), 1842–1866.

[44] Raja S Kushalnagar, Gary W Behm, Aaron W Kelstone, and Shareef Ali. 2015. Tracked speech-to-text display: Enhancing accessibility and readability of real-time speech-to-text. In *Proceedings of the 17th International ACM SIGACCESS Conference on Computers & Accessibility*. 223–230.

[45] Chao-Hsing Lee and Chien-Wen Chen. 2021. Impulse buying behaviors in live streaming commerce based on the stimulus-organism-response framework. *Information* 12, 6 (2021), 241.

[46] Mingwei Li, Qingjin Wang, and Ying Cao. 2022. Understanding consumer online impulse buying in live streaming e-commerce: A stimulus-organism-response framework. *International journal of environmental research and public health* 19, 7 (2022), 4378.

[47] Junxiang Liao, Zheng Wei, Zeyu Yang, Xian Xu, Pan Hui, Changyang He, and Muzhi Zhou. 2025. "Even When Success Seems Impossible, I Keep Streaming": How Do Chinese Elderly Streamers Interact with Platform Algorithmic (In)visibility. In *Proceedings of the 2025 CHI Conference on Human Factors in Computing Systems*. 1–15.

[48] Weiwei Lu, Yiwen Chen, and Shang Li. 2022. Mechanism of interaction and entertainment impact on impulse purchase intention in shopping livestream. In *Proceedings of the 6th International Conference on E-Commerce, E-Business and E-Government*. 59–63.

[49] Le Luo, Dongdong Weng, Guo Songrui, Jie Hao, and Ziqi Tu. 2022. Avatar interpreter: improving classroom experiences for deaf and hard-of-hearing people based on augmented reality. In *CHI Conference on Human Factors in Computing Systems Extended Abstracts*. 1–5.

[50] Emma J McDonnell, Ping Liu, Steven M Goodman, Raja Kushalnagar, Jon E Froehlich, and Leah Findlater. 2021. Social, environmental, and technical: Factors at play in the current use and future design of small-group captioning. *Proceedings of the ACM on Human-Computer Interaction* 5, CSCW2 (2021), 1–25.

[51] Patrick E McKnight and Julius Najab. 2010. Mann-Whitney U Test. *The Corsini encyclopedia of psychology* (2010), 1–1.

[52] J. Meinzen-Derr. 2023. Tech-Based Intervention Dramatically Improved Language Skills in Hearing Impaired Kids. https://www.physiciansweekly.com/tech-based-intervention-dramatically-improved-language-skills-in-hearing-impaired-kids/

[53] Mohammadreza Mirzaei, Peter Kan, and Hannes Kaufmann. 2020. EarVR: Using ear haptics in virtual reality for deaf and Hard-of-Hearing people. *IEEE transactions on visualization and computer graphics* 26, 5 (2020), 2084–2093.

[54] Meredith Ringel Morris, Kori Inkpen, and Gina Venolia. 2014. Remote shopping advice: enhancing in-store shopping with social technologies. In *Proceedings of the 17th ACM conference on Computer supported cooperative work & social computing*. 662–673.

[55] Surge News. 2024. China Live Streaming E-Commerce Industry Research Report 2023. https://www.thepaper.cn/newsDetail_forward_26518751

[56] Yi-Hao Peng, Ming-Wei Hsi, Paul Taele, Ting-Yu Lin, Po-En Lai, Leon Hsu, Tzu-chuan Chen, Te-Yen Wu, Yu-An Chen, Hsien-Hui Tang, et al. 2018. Speechbubbles: Enhancing captioning experiences for deaf and hard-of-hearing people in group conversations. In *Proceedings of the 2018 CHI Conference on Human Factors in Computing Systems*. 1–10.

[57] Michael V Perrotta, Thorhildur Asgeirsdottir, and David M Eagleman. 2021. Deciphering sounds through patterns of vibration on the skin. *Neuroscience* 458 (2021), 77–86.

[58] Anne Marie Piper and James D Hollan. 2008. Supporting medical conversations between deaf and hearing individuals with tabletop displays. In *Proceedings of the 2008 ACM conference on Computer supported cooperative work*. 147–156.

[59] Todd Andrew Ricketts. 2001. Directional hearing aids. *Trends in Amplification* 5, 4 (2001), 139–176.

[60] Mohamad Saifuddin Rusli and Zainuddin Ibrahim. 2022. Augmented Reality (AR) for Deaf and Hard of Hearing (DHH) for animation. *e-Academia Journal* 11, 2 (2022), 175–186.

[61] Rufat Rzayev, Paweł W Woźniak, Tilman Dingler, and Niels Henze. 2018. Reading on smart glasses: The effect of text position, presentation type and walking. In *Proceedings of the 2018 CHI conference on human factors in computing systems*. 1–9.

[62] Yasith Samaradivakara, Thavindu Ushan, Asela Pathirage, Prasanth Sasikumar, Kasun Karunanayaka, Chamath Keppitiyagama, and Suranga Nanayakkara. 2024. SeEar: Tailoring Real-time AR Caption Interfaces for Deaf and Hard-of-Hearing (DHH) Students in Specialized Educational Settings. In *Extended Abstracts of the CHI Conference on Human Factors in Computing Systems*. 1–8.

[63] Maxime E Sanders, Ellen Kant, Adriana L Smit, and Inge Stegeman. 2021. The effect of hearing aids on cognitive function: a systematic review. *PLoS One* 16, 12 (2021), e0261207.

[64] Matthew Seita, Sooyeon Lee, Sarah Andrew, Kristen Shinohara, and Matt Huenerfauth. 2022. Remotely co-designing features for communication applications using automatic captioning with deaf and hearing pairs. In *Proceedings of the 2022 CHI Conference on Human Factors in Computing Systems*. 1–13.

[65] Ellyn G Sheffield, Michael Starling, and Daniel Schwab. 2011. Bringing text display digital radio to consumers with hearing loss. *Journal of deaf studies and deaf education* 16, 4 (2011), 537–552.

[66] Ruo Si. 2021. *China livestreaming e-commerce industry insights.* Springer.

[67] Ruo Si and Ruo Si. 2021. The Evolvement of Livestreaming E-Commerce. *China Livestreaming E-commerce Industry Insights* (2021), 1–31.

[68] Preeti Srinivasan. 2020. " Am I Overwhelmed with this Information?" A Cross-Platform Study on Information Overload, Technostress, Well-Being, and Continued Social Media Usage Intentions. In *Companion Publication of the 2020 Conference on Computer Supported Cooperative Work and Social Computing*. 165–170.

[69] Anselm Strauss and Juliet Corbin. 1998. Basics of qualitative research techniques. (1998).

[70] Ningjing Tang, Lei Tao, Bo Wen, and Zhicong Lu. 2022. Dare to dream, dare to livestream: How e-commerce livestreaming empowers chinese rural women. In *Proceedings of the 2022 CHI conference on human factors in computing systems*. 1–13.

[71] Renu Thakur, Jaikishan Jayakumar, and Sangeeta Pant. 2019. A comparative study of visual attention in hearing impaired and normal schoolgoing children. *Indian Journal of Otology* 25, 4 (2019), 192–195.

[72] Renu Thakur, Jaikishan Jayakumar, and Sangeeta Pant. 2023. Visual Perception and Attentional Skills in School-age Children: A Cross-Sectional Study of Reading Proficiency in the Hearing Impaired. *Indian Journal of Community Medicine* 48, 4 (2023), 544–549.

[73] Gokhan Tur and Renato De Mori. 2011. *Spoken language understanding: Systems for extracting semantic information from speech.* John Wiley & Sons.

[74] Ye Wang, Zhicong Lu, Peng Cao, Jingyi Chu, Haonan Wang, and Roger Wattenhofer. 2022. How live streaming changes shopping decisions in E-commerce: A study of live streaming commerce. *Computer Supported Cooperative Work (CSCW)* 31, 4 (2022), 701–729.

[75] Diana Teresia Spits Warnars, Leonardo Primadala Putra, Harco Leslie Hendric Spits Warnars, Wiranto Herry Utomo, et al. 2019. Intelligent E-commerce for Special Needs. In *2019 7th International Conference on Cyber and IT Service Management (CITSM)*, Vol. 7. IEEE, 1–5.

[76] Zheng Wei, Yuzheng Chen, Wai Tong, Xuan Zong, Huamin Qu, Xian Xu, and Lik-Hang Lee. 2024. Hearing the Moment with MetaEcho! From Physical to Virtual in Synchronized Sound Recording. In *Proceedings of the 32nd ACM International Conference on Multimedia*. 6520–6529.

[77] Zheng Wei, Lik-Hang Lee, Wai Tong, Xian Xu, Chaozhe Zhang, Huamin Qu, and Pan Hui. 2025. "I Can't Even Recall What I Bought": How Design Influences Impulsive Buying in Douyin Live Sales. *International Journal of Human–Computer Interaction* (2025), 1–18.

[78] Zheng Wei, Junxiang Liao, Lik-Hang Lee, Huamin Qu, and Xian Xu. 2025. Towards Enhanced Learning through Presence: A Systematic Review of Presence in Virtual Reality Across Tasks and Disciplines. *arXiv preprint arXiv:2504.13845* (2025).

[79] Zheng Wei, Jia Sun, Junxiang Liao, Lik-Hang Lee, Chan In Sio, Pan Hui, Huamin Qu, Wai Tong, and Xian Xu. 2025. Illuminating the scene: How virtual environments and learning modes shape film lighting mastery in virtual reality. *IEEE Transactions on Visualization and Computer Graphics* (2025).

[80] Zheng Wei, Xian Xu, Lik-Hang Lee, Wai Tong, Huamin Qu, and Pan Hui. 2023. Feeling Present! From Physical to Virtual Cinematography Lighting Education with Metashadow. In *Proceedings of the 31st ACM International Conference on Multimedia*. 1127–1136.

[81] Thomas Westin, José Neves, Peter Mozelius, Carla Sousa, and Lara Mantovan. 2022. Inclusive AR-games for education of deaf children: Challenges and opportunities. In *European Conference on Games Based Learning*, Vol. 16. 597–604.

[82] Sumas Wongsunopparat and Binmei Deng. 2021. Factors influencing purchase decision of Chinese consumer under live streaming E-commerce model. *Journal of Small Business and Entrepreneurship* 9, 2 (2021), 1–15.

[83] Qunfang Wu, Yisi Sang, Dakuo Wang, and Zhicong Lu. 2023. Malicious selling strategies in livestream E-commerce: A case study of alibaba's taobao and ByteDance's TikTok. *ACM Transactions on Computer-Human Interaction* 30, 3 (2023), 1–29.

[84] Xian Xu, Wai Tong, Zheng Wei, Meng Xia, Lik-Hang Lee, and Huamin Qu. 2023. Cinematography in the metaverse: Exploring the lighting education on a soundstage. In *2023 IEEE conference on virtual reality and 3d user interfaces abstracts and workshops (VRW)*. IEEE, 571–572.

[85] Xian Xu, Wai Tong, Zheng Wei, Meng Xia, Lik-Hang Lee, and Huamin Qu. 2024. Transforming cinematography lighting education in the metaverse. *Visual Informatics* (2024).

[86] Fu-Chia Yang. 2022. *Holographic Sign Language Interpreter: A User Interaction Study within Mixed Reality Classroom.* Master's thesis. Purdue University.

[87] Chongjae Yoo and Hwanhee Lee. 2023. Improving Abstractive Dialogue Summarization Using Keyword Extraction. *Applied Sciences* 13, 17 (2023), 9771.

[88] Jeewoo Yun, Don Lee, Michael Cottingham, and Hyowon Hyun. 2023. New generation commerce: The rise of live commerce (L-commerce). *Journal of Retailing and Consumer Services* 74 (2023), 103394.

# A APPENDIX

### Table A1. System Experience Questionnaire

| Dimensions | Ranging from 1 (Strongly Disagree) to 7 (Strongly Agree) |
|---|---|
| Usability | This system is very helpful for me when watching livestream shopping. |
| | Using this system, my livestream shopping experience has become easier. |
| Ease of Use | I find this system's interface user-friendly and the functions easy to use. |
| | I can quickly get started and operate this system to watch livestream shopping. |
| Content Comprehensibility | Using this system makes it easier for me to understand the livestream content. |
| | I can better keep up with the pace of the facilitator through this system. |
| Content memorability | This system allows me to more easily remember important livestream information. |
| | Using this system, I can clearly recall what the facilitator discussed. |
| Accuracy of Summaries | This system is very accurate in summarizing the content of livestream shopping. |
| | This system can help me quickly grasp the core information during the livestream shopping. |
| Viewing interest | This system makes me feel more interested in watching livestream shopping. |
| | After using this system, I will be more inclined to watch livestream shopping. |

### Table A2. NASA-TLX Dimensions and Corresponding Questionaire

| Dimension | Please mark your rating on the scale line for each dimension from 0 (lowest) to 100 (highest). |
|---|---|
| Mental Demand | How much mental and thought activity do you feel is required when watching a live tape? |
| | How difficult was the task and did it require complex thinking and decision-making? |
| Physical Demand | How much physical activity do you feel is required while watching the live tape video? |
| | How physically taxing was the task? |
| Temporal Demand | How much time pressure is there when watching live tape recordings? |
| | Did you feel like you had plenty of time, or were you pressed for time? |
| Performance | How do you feel you performed on the task? |
| | Were you able to use our software effectively to complete the task? |
| | (reverse scoring, i.e. lower scores indicate better performance) |
| Effort | How much effort do you think was put in to effectively use our software to watch live band videos? |
| | Did the task require a lot of effort on your part? |
| | (Effort is the total amount of physical and mental effort you need to put in to complete the task. |
| | It measures the amount of effort and resources you need to put into the task in order to complete it |
| | while using the software to watch live tape recordings.) |
| Frustration | To what extent do you feel frustrated, disheartened or unhappy when using the software to watch live tape recordings? |
| | Did you find the task annoying or satisfying? (Frustration is the degree of frustration, disappointment, |
| | and unpleasantness you feel in completing a task. It measures the difficulties and obstacles you encountered |
| | while using the software to watch the live tape recordings and the negative emotional impact these difficulties had on you.) |

## A.1 Accuracy Measuring

*A.1.1 Criteria.* Record the screen on selected livestream on a mobile device, play the recordings on the test device, and activate the application function. In five minutes, the program evaluates the prerecorded video every 30 seconds, for a total of 10 evaluations. During the evaluation, the evaluator also manually records the information. This information is compared with manually recorded key information (Including promotional policies, free shipping availability, return policy, pricing, after-sales service, product introduction, usage experience and user manual). If none of the following situations (A4) occurs, score 1 point; otherwise, no points are awarded. The maximum score is 10 points.

For Condensed Text Display, because its transaction word extraction occurs over a short time span, you should focus typically on whether it correctly identifies transaction, particularly the accuracy of names and numbers. For example, it might incorrectly recognize word ' 垫子'(mat) as ' 电池'(battery).

In the Summary Framework Display, due to its longer time span and more complex logical relationships, the evaluation emphasizes the program's ability to handle the scenarios, such as determining whether the nested relationships between product categories and subcategories are correctly identified, understanding the relationship between original prices and discounted prices, recognizing newly introduced promotions during the live-stream, and distinguishing between different products.

Table A3.  **Live Information Memory Test**

| Demand | Number | Question | Answer |
|---|---|---|---|
| Product Features | 1 | What are the product features of this latex cooler? | Latex is synthetic; formaldehyde-free; meets Latex Adult Class B standards; double-thick finish. |
| | 2 | What are the product features of this toothpaste? | Fluorinated, potassium nitrate formulations; non-fluorinated, potassium nitrate formulations; fluorinated, potassium chlorate formulations; non-fluorinated, potassium chlorate formulations. |
| Discount Policies | 3 | What is the gift policy of this latex cooler? | The live video doesn't say; buy one get one free; buy two get one free; buy three get one free. |
| | 4 | What is the gift policy of this toothpaste? | Get 2 toothbrushes + 2 toothpastes; Get 2 toothbrushes + 1 toothpaste; Get 7 toothbrushes + 2 toothpastes; Get 6 toothbrushes + 2 toothpastes; |
| Price Changes | 5 | What is the preferential policy of this latex mat? | Today's special is 99-140+ yuan; Today's special is 69-149 yuan; Today's special is 99-149 yuan; Today's special is 69-140+ yuan. |
| | 6 | What is the preferential policy of this toothpaste? | 5 toothpastes for 79 yuan; 5 toothpastes for 191 yuan; 5 toothpastes and giveaway for 79 yuan; 5 toothpastes and giveaway for 191 yuan. |
| 7-day No Reason Return Policy | 7 | What is the return policy of this latex mat? | Partially support 7 days return without reason; Only support 7 days return - without reason for quality problems; Unconditionally support 7 days - return without reason for any conditions; Do not support 7 days return without reason. |
| | 8 | What is the return policy of this toothpaste? | Do not support 7 days no reason to return; only support quality problems 7 days no reason to return; unconditional support for any conditions of 7 days - no reason to return; livestreamer did not say |
| Product Functionality Introduction | 9 | How many specifications does this latex mat have? | 1 type, 1.2m single bed for dormitory use; 2 types, 1.2m and 1.5m; 3 types, 1.2m and 1.5m and 1.8m; 4 types, 1.2m and 1.5m and 1.8m and 2.4m. |
| | 10 | How many grams of this toothpaste? | 490g, lasts 1 month; 590g, lasts 3 months; 490g, lasts 3 months; 590g, lasts 1 month. |

Table A4.  Explanation of Discrepancy Reasons

| Discrepancy Reason | Explanation |
|---|---|
| Incorrect Information | Displayed information does not match the video content |
| Confused Information | New and old product information is mixed |
| Missing Information | Key information is missing |
| Fabricated Information | Displayed information is not given in the video |
| Interpret Out of Context | Displayed information is partially correct |

*A.1.2 Examples.* During a live-stream saling, the host introduces a pure cotton T-shirt originally priced at 59 CNY, now discounted to 9.9 CNY. The product includes free shipping and offers unconditional return or exchange. The host claims that its quality is comparable to T-shirts priced at several dozen yuan. Here are some cases about discrepancy:

(1) **Incorrect Information:** The price is 5.9 CNY. (The true value should be 9.9 CNY or 59 for the original price)
(2) **Confused Information:** *Product:* A several dozen yuan T-shirt (Mix it with another T-shirt.)
(3) **Fabricated Information:** *Product:* Wireless Bluetooth Earbuds (This usually happens at the very beginning. If the host does not provide any product information, due to model hallucination, it may instead introduce a sample product.)
(4) **Interpret Out of Context:** *Product:* Red-color Pure Cotton T-shirt (The *red* one is only a subcategory of the T-shirt.)

A correct framework output as follows:

**Product:** Pure Cotton T-Shirt
**Category:** Clothing
**Promotion Policy:** Original price 59 CNY, now 9.9 CNY
**Free Shipping:** Yes
**7-day Unconditional Return:** Yes
**Price:** 9.9 CNY
**After-Sales Service:** Full refund or exchange available
**Product Description:** High-quality pure cotton T-shirt, like a dozens CNY one
**Usage Experience:** Comfortable, breathable, and durable fabric for everyday wear
**User Manual:** Wash in cold water, do not bleach, and air dry for best results

## A.2 Interactions with LLM

There are three instances of LLM interactions in the *EChoAid* operation process: Condensed Text Display, Emoji, and Summary Framework Display. The APIs interacting with ERNIE include:

- **System Prompt:** Optional. The background information and instructions.
- **User Input:** The main content of the question posed.
- **LLM Response:** Response from ERNIE's server.

*EchoAid* interacts with the *ERNIE* APIs by HTTP requests from the mobile through a dialogue format. ERNIE supports follow-up questions based on previous inputs and outputs to further develop the response. In the *Emoji* task, this method is utilized, so the dialogue alternates between the LLM response and the User input.

*A.2.1 Condensed Text Display.* **Dialogue:**

(1) **System Prompt:** *Prompt1* (The prompt shown below)
(2) **User Input:** *Raw Transcription in 40 secs* (Dialogues may contain unclear and interfering parts. Choosing a 40-second time window instead of a 30-second one provides more speech content, smoothing transitions between different rounds when handling information at the boundaries. This helps correct errors and prevent information loss, thereby enhancing the program's robustness.)
(3) **LLM Response:** *Output1*

**Prompt1 (Chinese):**
能力与角色: 你是一个直播带货内容处理助手，帮助听障人士客户了解商品信息。
背景信息: 你的原始输入是直播带货主播的语音转文字内容。你的任务是接收并浓缩输入的录音转文字内容。注意，语音转录内容可能有重复、遗漏、识别错误的情况，特别是靠近

末尾 10 秒长度的字符，因为它尚未在语音模型纠正。你需要仔细分析上下文，理解主播的信息。如果无法理解，请忽略这些信息。

指令：你的任务是接收并浓缩输入的录音转文字内容，输出不超过 50 字。提取最关键的信息。主播会有重复对话，这些应该忽略。仅提取文件中的信息，不需要扩写。

输出风格：请你以主播的口吻输出。

下面是输入：

**Prompt1 (English):**

Role and Capability: You are a live-stream e-commerce content assistant, helping hard of hearing customers understand product information.

Background Information: Your input is the live-stream host's speech converted into text. Your task is to condense the transcription into concise content. Note that the transcription may include repetition, omissions, or recognition errors, especially in the last 10 seconds of text due to the speech model's corrections still being applied. You need to carefully analyze the context and understand the host's information. If the content is unintelligible, ignore it.

Instruction: Your task is to condense the transcription into a maximum of 50 words. Extract only the most critical information. Ignore repeated dialogue. Only summarize information from the text; do not expand on it.

Output Style: Write in the tone of the host.

Here is the input:

*A.2.2  Emoji.* **Dialogue:**

(1) **System Prompt:** *Prompt1* (The prompt in A.2.1).
(2) **User Input:** *Raw Transcription in 40 secs*
(3) **LLM Response:** *Output1* (**Response in A.2.1**)
(4) **User Input:** *Prompt2* (The prompt below)
(5) **LLM Response:** *Output2*

**Prompt2 (Chinese):** 下面是相关的 Emoji。找到和主播内容相关的 Emoji。

👍: 主播特别推荐该产品。
👎: 主播不推荐该产品。
⭐: 用于表示产品评级，可以根据星级数量显示产品质量。
⏱: 表示有较短限时（5 分钟内）的活动（如抢购、抽奖）正在进行。
⌛: 强调时间紧迫，即将结束。
✨: 用于强调产品的特殊功能或亮点。
🔍: 表示主播正在详细展示产品细节。
🎁: 表示主播正在介绍新产品。
🆕: 强调产品是新上市的。
💲: 主播正在介绍价格。
📢: 表示有促销政策。
🎟: 表示有优惠券或折扣可用。
🏷: 显示具体的折扣价。
👋: 表示直播即将结束，主播正在致谢。
☎: 表示感谢观众参与。

输出时，列举所有符合要求的 Emoji。有强调时，可以输出复数个 Emoji。解释理由时，直接解释，不输出 Emoji。同时，不需要总结。

**Prompt2 (English):**

Below are the relevant emojis. Find the emojis relevant to the host's content.

👍: The host particularly recommends the product.
👎: The host does not recommend the product.

✩: Represents product ratings, with the number of stars indicating quality.
⏱: Indicates a short-term event (within 5 minutes) such as a flash sale or giveaway.
⧖: Emphasizes urgency, indicating the offer is about to end.
✧: Highlights special features or unique aspects of the product.
🔍: Indicates the host is providing detailed product demonstrations.
🎁: Indicates the host is introducing a new product.
🆕: Emphasizes the product is newly launched.
💲: Indicates the host is discussing pricing.
🏷: Highlights promotional offers.
🎟: Indicates available coupons or discounts.
🧾: Displays specific discounted prices.
👋: Signals the live stream is ending, and the host is thanking the audience.
🙏: Expresses gratitude to the viewers for participating.

When outputting, list all relevant emojis. Use multiple emojis when emphasize. For explanations, provide the reasoning directly, without using emojis. Summarization is not required.

### A.2.3 Summary Framework Display. **Dialogue:**

(1) **System Prompt:** *Prompt3* (The prompt below.)
(2) **User Input:** *Up to 10 responses in A.2.1 concat Prompt3*
(3) **LLM Response:** *Output3*

**Prompt3 (Chinese):** 你是一个直播带货内容处理助手，帮助听障人士客户了解商品信息。要求如下：

0. 你的原始输入是经过总结的直播带货主播的语音转文字内容。你的输入会包含最多 10 条内容，按时间升序排序，每条间隔 30 秒。

1. 你的任务是接收并提取关键词。以以下格式输出（未提及的部分，请用 null 代替）：

商品: ...
类别: ...
促销政策: ...
是否包邮: ...
7 天无理由退货: ...
价格: ...
售后服务: ...
产品介绍: ...
使用体验: ...
使用说明书: ...

2. 注意，你必须严格按格式输出。不需要任何解释，禁止输出头尾的 "```"。一旦格式解析错误，你会被立刻杀死。

3. 语音转录内容可能有重复、遗漏、识别错误的情况，特别是靠近末尾 10 秒长度的字符，因为它尚未在语音模型纠正。你需要仔细分析上下文，理解主播的信息。

4. 提取最关键的信息。主播会有很多故事、语气词、无效对话，这些应该忽略。

5. 如果有信息，你需要以主播的视角浓缩文字。

6. 浓缩的文字一段一句，语言简练，只保留最关键信息。

7. 记住，输出我要求的，不需要输出 "```"。否则你会被杀死的。这无法解析。

开始处理。

**Prompt3 (English):**
You are a live-stream e-commerce content assistant helping hard of hearing customers understand product information. The requirements are as follows:

0. Your input is summarized text converted from the host's speech. It contains up to 10 items sorted in ascending time order, with each item spaced 30 seconds apart.

1. Your task is to extract keywords and output in the following format (use 'null' for unspecified fields):

Product: ...
Category: ...
Promotional Policy: ...
Free Shipping: ...
7-Day No Reason Return: ...
Price: ...
After-Sales Service: ...
Product Description: ...
User Experience: ...
User Manual: ...

2. You must strictly adhere to the format. Do not provide any explanations, and do not include the opening or closing "```". Failure to parse the format correctly will result in immediate termination.

3. Transcription may contain repetitions, omissions, or recognition errors, especially near the last 10 seconds of text due to incomplete corrections by the speech model. You need to analyze the context carefully to understand the host's message.

4. Extract the most critical information. Ignore irrelevant details such as stories, filler words, or casual dialogue.

5. If there is useful information, condense it from the host's perspective.

6. Use concise, impactful sentences with only the most essential information.

7. Remember to provide the required output without enclosing it in "```". Failure to comply will result in termination. It is unable to resolve.

Start processing.

## A.3 Live Streaming Platform Experience Questionnaire

(1) For you, how often do you choose to watch live streams? If you rarely or do not watch live streams, can you share what factors influenced your decision?

(2) what type of live content do you typically prefer to watch? What factors determine this preference?

(3) Have you ever purchased a product or service while watching a live stream? Can you share what prompted you to make a purchase decision?

(4) Do you feel that existing live streaming platforms take into account the needs of the hearing impaired? What is your viewing experience on these platforms?

(5) When watching live streams, do you feel overwhelmed by the amount of information and find it difficult to process? Did this feeling affect your viewing experience?

(6) Which social media platforms do you use to communicate on a daily basis? What are the main reasons for choosing these platforms?

(7) Do you participate in interactions with the anchor or other viewers while watching the live broadcast? What factors affect your willingness to interact?

(8) Have you ever tried to communicate with the anchor or other viewers? If you have not yet tried, what has prevented you from doing so?

(9) In the course of watching the live broadcast, were there any technical limitations that made you feel that your identity and needs were not fully recognized? Can you describe which technical limitations?

(10) What improvements do you think are needed for the live streaming platform to better serve the hearing impaired?

## A.4 Live Streaming Platform Experience Questionnaire

(1) What features appeal to you most when using EchoAid? Why?

(2) What do you think are the most outstanding features of EchoAid?

(3) Is there anything that bothers or inconveniences you in the process of using it? Which ones specifically?

(4) Are there features that could be optimized or improved in your opinion? Expand on that a bit?

(5) How useful do you find EchoAid in helping you get live shopping information? Can you give specific examples?

(6) Has EchoAid been effective in reducing information overload? What changes can you feel?

(7) Did you find the interface of EchoAid easy to use?What was smooth or complicated to operate? Did you have to spend time learning the features when first using it?

(8) How do you feel EchoAid has positively impacted your daily life or shopping experience?